\documentclass[a4paper,fleqn,usenatbib]{mnras}

\usepackage{newtxtext,newtxmath}
\usepackage[T1]{fontenc}
\usepackage{ae,aecompl}

\usepackage{graphicx}	% Including figure files
\usepackage{amsmath}	% Advanced maths commands
\usepackage{amssymb}	% Extra maths symbols
\usepackage[usenames]{color}

\newcommand{\fornaxTwoDbetazero}{-0.85 \pm 0.06}
\newcommand{\fornaxTwoDbetaone}{1.52^{+0.36}_{-0.48}}

\newcommand{\fornaxThreeDbetazero}{-1.07 \pm 0.17}
\newcommand{\fornaxThreeDbetaone}{1.56^{+0.36}_{-0.46}}

\newcommand{\chimerabetazero}{-0.52^{+0.19}_{-0.22}}
\newcommand{\chimerabetaone}{1.45^{+0.40}_{-0.38}}

\newcommand{\coconutbetazero}{-0.73 \pm 0.11}
\newcommand{\coconutbetaone}{1.49^{+0.46}_{-0.35}}

\newcommand{\deltafornaxTwoD}{0.73 \pm 0.05}
\newcommand{\deltafornaxThreeD}{0.95 \pm 0.06}
\newcommand{\deltachimera}{0.33 \pm 0.06}
\newcommand{\deltacoconut}{0.62 \pm 0.05}
\newcommand{\muobs}{-0.13 \pm 0.05}
\newcommand{\sigmaobs}{0.21^{+0.05}_{-0.04}}

\title[]{A Comparison of Explosion Energies for Simulated and Observed Core-Collapse Supernovae}

\author[Murphy et al.]{Jeremiah W. Murphy$^{1}$\thanks{E-mail:
    jwmurphy@fsu.edu}, Quintin Mabanta$^{1}$, Joshua C. Dolence$^2$.
\\$^1$Department of Physics, Florida State University, 77 Chieftan
  Way, Tallahassee, FL 32306, USA
\\$^2$Los Alamos National Laboratory}

\date{Accepted XXX. Received YYY; in original form ZZZ}

\pubyear{2019}

\begin{document}
\label{firstpage}
\pagerange{\pageref{firstpage}--\pageref{lastpage}}
\maketitle

\begin{abstract}
There are now $\sim$20 multi-dimensional core-collapse supernova
(CCSN) simulations that explode. However, these simulations have explosion
energies that are a few times $10^{50}$ erg, not $10^{51}$
erg.  In this
manuscript, we compare the inferred explosion energies of these simulations
and observations of 38 SN~IIP.  Assuming a log-normal distribution, the mean explosion energy for the
observations is
$\mu_{\rm obs} =  -0.13\pm 0.05$ 
($\log_{10}(E/10^{51}\, {\rm erg})$) and the width is
$\sigma_{\rm obs} = 0.21^{+0.05}_{-0.04}$.
Only three CCSN codes have sufficient simulations to
compare with observations: CHIMERA, CoCoNuT-FMT, and FORNAX.
Currently, FORNAX has the largest sample of simulations.
The two-dimensional FORNAX simulations show a correlation between
explosion energy and progenitor mass, ranging from linear to
quadratic, 
$E_{\rm sim} \propto M^{1-2}$; this correlation is consistent with inferences from observations.  In addition, we infer
the ratio of the observed-to-simulated explosion energies, 
$\Delta=\log_{10}(E_{\rm obs}/E_{\rm sim})$.  
For the CHIMERA set, $\Delta=0.33\pm0.06$;
for CoCoNuT-FMT, $\Delta=0.62\pm0.05$;
for FORNAX2D, $\Delta=0.73\pm0.05$, 
and for FORNAX3D, $\Delta=0.95\pm0.06$.  
On average, the simulations are less energetic than inferred energies from
observations ($\Delta \approx 0.7$), but we also note that the variation among the
simulations (max($\Delta$)-min($\Delta$) $\approx 0.6$) is as large as
this average offset.  This suggests that further improvements to the
simulations could resolve the discrepancy.  Furthermore, both the
simulations and the observations are heavily biased. In this preliminary
comparison, we model these biases, but to more reliably compare the
explosion energies, we recommend strategies to unbias both the
simulations and observations.
\end{abstract}


%%%%%%%%%%%%%%%%%%%%%%%%%%%%%%%%%%%%%%%%%%%%%%%%%%

%%%%%%%%%%%%%%%%% BODY OF PAPER %%%%%%%%%%%%%%%%%%

\section{Introduction}\label{sec:intro}

A primary goal of core-collapse supernovae theory is to predict which
stars will explode, but for more than two decades, the more pressing
challenge has been to produce at least one successful explosion in
numerical simulations.  Recent multi-dimensional simulations are
finally producing self-consistent explosions \citep{lentz2015,mueller2015,bruenn2016,melson2015a,summa2016,radice2017,oconnor2018,ott2018,vartanyan2018a,mueller2019,vartanyan2019,burrows2019}.  While there are still
only a handful of simulations with successful explosions, a trend is
already emerging; the explosion energies of simulations tend to be
less energetic than explosion energies inferred from observations.  In this manuscript, we
quantify the discrepancy between simulations and observations.

Over the last several decades, CCSN simulations have become much more
computationally expensive (requiring 10s of millions of CPU-hours) but they also seem to be converging toward
successful explosions.  \citet{colgate1966} was the first to suggest
that the change in gravitational energy due to core collapse could power the
supernova explosion; they also suggested that neutrinos transfer this
energy from the core to the mantle.  However, more
detailed modeling indicates that the bounce shock quickly stalls into
an accretion shock due to electron capture and neutrino losses but
mostly due to nuclear disassociation \citep{hillebrandt1981,mazurek1982}.  Using one-dimensional neutrino
radiation hydrodynamic simulations, \citet{wilson85} and
\citet{bethe85} suggested that neutrinos
eventually relaunch the stalled shock into an explosion.  However,
more modern one-dimensional simulations show that most stars do not
explode.  During the 1990s, two-dimensional simulations using gray
flux-limited diffusion hinted that convection might
aide the explosion when 1D failed \citep{benz94,herant94,burrows95,janka95}.  \citet{murphy08b} investigated
the conditions for explosion and found that the neutrino luminosity
required for explosion is 30\% less in 2D than 1D.  \citet{mabanta2018}
derived the conditions for explosion with and without a convection
model.  They found that the convection model does reduce the explosion
condition by 30\% in agreement with simulations, and they found that a
large part of the reduction is caused by turbulent dissipation.

These investigations suggest a minimum set of requirements for
self-consistent core-collapse supernova simulations.  General
relativity (GR) is likely important, so the code should employ GR or
at least a post Newtonian potential inspired by GR.  Neutrino
transport should include the interactions for electron, mu, and tau
flavors, and it should be multi-angle and multi-energy.  Finally, the
simulations should be multi-dimensional, preferably
three-dimensional, but two-dimensional simulations have shown similar
explosion conditions \citep{hanke12} and energetics \citep{burrows2019}.
Following is a
list of publications that report explosive simulations using codes
with these minimum requirements: \citep{lentz2015,mueller2015,bruenn2016,melson2015a,summa2016,radice2017,oconnor2018,ott2018,vartanyan2018a,mueller2019,vartanyan2019,burrows2019}.  Of these, the following
publications report positive explosion energies that begin to plateau
in energy:
\citep{mueller2015,bruenn2016,melson2015a,radice2017,vartanyan2018a,mueller2019,vartanyan2019,burrows2019}.
In general, the explosion energies reported range from $\sim$0.1
to $\sim 0.9 \times 10^{51}$ erg.

One may infer explosion energies of observed type IIP SNe (SN~IIP) by modeling the
light curve and spectra of type IIP SNe (SN~IIP) \citep{arnett1980}.
The velocity, brightness, and duration of the plateau depend mostly upon the explosion energy,
ejecta mass, nickel mass, and progenitor radius \citep{popov1993,kasen2009,dessart2019,goldberg2019}.  Therefore, given spectra and light curves,
one may infer the explosion energy.  There are two
general techniques to perform this inference.  One is to force
explosions in one-dimensional simulations and model
the expansion, spectra, and light curve
\citep{kasen2009,sukhbold2016,goldberg2019}.  In the rest of this
manuscript, we refer to this as photospheric modeling.  The other is to use fitting
formulae to connect the photospheric parameters to model parameters \citep{arnett1980,chugai1991,popov1993}.
Generally, photospheric modeling is used to calibrate the fitting
formulae.  A recent investigation of photospheric modeling that also
inferred explosion energies was
performed by \citet{kasen2009}.  They confirm the previous fitting formulae
of \citet{popov1993} and very roughly infer explosion
energies for SN~IIP that range from 0.5 to $4.0 \times 10^{51}$ erg.

While promising, photospheric modeling of SN~IIP presents significant
challenges.  The most recent investigations indicate degeneracies when
inferring nickel mass, ejecta mass, explosion energy, and progenitor radius
\citep{dessart2019,goldberg2019}.  In fact, \citet{goldberg2019}
suggest that extra information on one of these parameters is required
to break the degeneracies.

\citet{pejcha2015} explored how these degeneracies affect the
statistical inference of these parameters.  They and \citet{muellerprieto2017} use the fitting formulae and statistical inference to not only infer the most likely explosion
energy but also the uncertainty and covariances associated with each inference.  Since
they use the fitting formulae for their inference, \citet{pejcha2015} caution that the inferred
explosion energies may not be as precise as using the light-curve
models.  Instead, they suggest that their study provides a systematic
investigation of the uncertainties and correlations in the inference.
Later, in section \ref{sec:observations}, we demonstrate that the distribution of
explosion energies from the light-curve modeling and fitting formulae
are similar.  Despite the concerns, this suggests that inferred explosion energies in
\citet{pejcha2015} may actually be as precise as the photospheric modeling
results of \citet{kasen2009}.  Since the results of \citet{pejcha2015}
and \citet{muellerprieto2017}
also provide inferred uncertainties, we propose using their results to represent the inferred explosion energies for observations.

Again the explosion energies of multi-dimensional simulations range
from 0.1 to $0.9 \times 10^{51}$ erg, yet the inferred explosion energies of observations range from
0.5 to $4.0 \times 10^{51}$ erg.  This suggests that the current set
of 
multi-dimensional simulations are less energetic than the energies
inferred from observations.  In this manuscript, we perform a preliminary comparison of the
explosion energies of observations and multi-dimensional
simulations.  In section \ref{sec:observations}, we discuss the observations, compare
explosion energies inferred by photospheric modeling and formula
fitting, and characterize the distribution of observed explosion
energies.  In section \ref{sec:simulations}, we describe the sample of multi-dimensional
simulations.  All of the simulations show a trend toward an asymptotic
explosion energy, but only a few actually reach an asymptotic
explosion energy.  Therefore, in section \ref{sec:extrapolating}, we propose a model for
the asymptotic explosion energy and infer an extrapolated explosion
energy for each simulation.  Each set of simulations does not yet sample the
full range of progenitors that lead to SN~IIP.  Therefore, in section
\ref{sec:InferEsimDistribution}, we assume a model correlating explosion energy and progenitor
mass, infer the parameters of this model, and use the results to infer
the full distribution of simulation explosion energies.  Then we
compare the simulations and observations in section \ref{sec:compare} and infer the
discrepancy between them.  Finally, in section \ref{sec:Conclusion}, we summarize
and discuss how to improve the inference by addressing biases in both
simulations and observations.

\section{Inferred Explosion Energies from Observations}
\label{sec:observations}

For the inferred explosion energies of observations, we consider three
sources: \citet{kasen2009}, \citet{pejcha2015}, and
\citet{muellerprieto2017}.  \citet{kasen2009} use both light curve and
spectra modeling to infer explosion energies from observations.  There are more recent photospheric
modeling papers \citep{sukhbold2016,dessart2019,morozova2018,goldberg2019}.
However, three of these do not actually infer explosion energies that
are based upon observations;
their focus is on providing better models and understanding the
degeneracies \citep{sukhbold2016,dessart2019,goldberg2019}.
\citet{morozova2018} do attempt to infer explosion energies.  However,
their inference is based upon light curve modeling alone, and does not
include the important velocity constrains provided by spectra.  Given
that there are significant degeneracies among the model parameters, it
is important to include all observational constraints
\citep{dessart2019,goldberg2019}.  The other two sources that we consider, \citet{pejcha2015} and \citet{muellerprieto2017}, use
statistical inference methods to
infer the explosion energy, nickel mass, and ejecta mass for 38 Type
II-Plateau supernovae.  See Table~\ref{tab:eobstable} for the list of
SNe.
%The sample includes SN~1980K, SN~1992H, SN~1995ad,
%SN~1996W, SN~1999em, SN~2001dc, SN~2004A, SN~2004dj, SN~2004et,
%SN~2005cs, SN~2008bk, SN~2008in, SN~2009N, SN~2009bw, SN~2009dd,
%SN~2009js, SN~2012A, SN~2012aw, SN~2013am.  
To infer these model
parameters, they used $M_V$, the absolute V band magnitude, $t_p$, the duration of the optically thick plateau
phase, and $v$, the expansion velocity of the photosphere.  Both
$M_{V}$ and $v$ were evaluated at the midpoint of the plateau phase.
In their likelihood model, the explosion parameters are related to
these observations via analytic scalings that were calibrated using
one-dimensional radiation-hydrodynamic explosion models.  These models
do not simulate the explosions self-consistently.  Rather, they force
the explosions, and relate the simulated light curve and expansion
velocity to the forced explosion energy, ejecta mass, and nickel
mass.  

To date, the inferences of \citet{pejcha2015} and
\citet{muellerprieto2017} represent the most
thorough analysis of the uncertainty and covariances for the
explosion parameters.  On the other hand, \citet{pejcha2015} also
caution that the mode of their inferences may not be as reliable as
inferences based upon numerical modeling of light curves.  However, as
yet, there are no statistical inferences using numerical modeling.
Therefore, to assess the accuracy of the \citet{pejcha2015} and
\citet{muellerprieto2017} statistical
inferences, we compare their distribution of explosion energies with
the distribution from \citet{kasen2009}, who model the light curve and
expansion velocities.  \citet{kasen2009} do not present a formal
inference.  However, in the right panel of their Figure~16, they
compare explosion energy as a function of photospheric velocity and
$M_V$.  On the same figure, they present observations for 22 SNe.
From this figure, we create a fitting function for the explosion
energy and use this to simply infer the explosion energy for each SN.
Note that this is not a thorough statistical inference, just simple
inference using a fitting formula.

\begin{figure}
%	% Allowable file formats are eps or ps if compiling using latex
%	% or pdf, png, jpg if compiling using pdflatex
	%\includegraphics[width=\columnwidth]{data/ComparePejchaKasen.pdf}
	\includegraphics[width=\columnwidth]{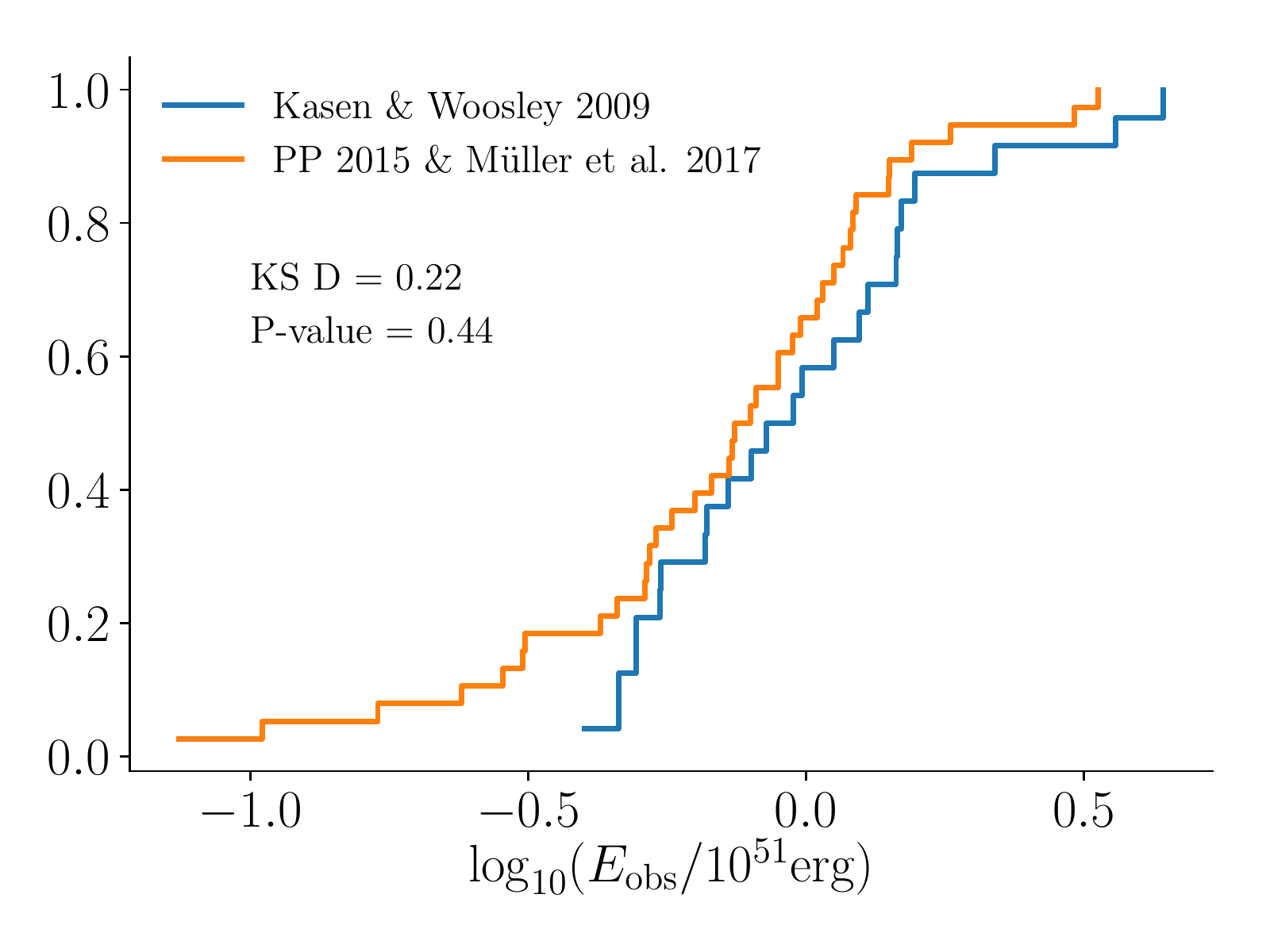}
    \caption{Comparing $E_{\rm obs}$, inferred explosion energies from
	observations.  \citet{kasen2009} modeled the light curve and
	spectrum of CCSNe and compared the modeled light curves and spectra
	with observations.  The blue curve represents their inference.
	\citet{pejcha2015} and \citet{muellerprieto2017} used analytic scalings to model the light curve
	and spectrum.  These scalings were calibrated using models similar
	to those in \citet{kasen2009}.  The primary purpose of
	\citet{pejcha2015} was to do a detailed statistical inference to
	better infer uncertainties and correlations among the parameters.
	The orange line shows the most likely $E_{\rm obs}$ for their
	analysis.  The KS test suggests that the distributions are
	consistent. Since the inferred $E_{\rm obs}$ from
	\citet{pejcha2015} and \citet{muellerprieto2017} have uncertainty estimates, we use their set
	when comparing to $E_{\rm sim}$.}
    \label{fig:Eobs}
\end{figure}

Figure~\ref{fig:Eobs} compares the distributions of explosion energies
for the two samples: one based upon photospheric modeling
\citep{kasen2009}; the other is based upon fitting formulae \citep{pejcha2015,muellerprieto2017}.  The KS test gives a D statistic of 0.22 and a
P-value of 0.44.  The two distributions are consistent with being
drawn from the same distribution.  
Even though \citet{pejcha2015} suggest that the precision
of their inference may not be as good as the simulations of
\citet{kasen2009}, we find that they are actually in agreement.
Since the \citet{pejcha2015} and \citet{muellerprieto2017} inference
provides both an explosion energy and an uncertainty, we use their
inferred explosion energies for the observational set.

Figure~2 of \citet{pejcha2015} gives their posterior distributions for
Ni mass and explosion energy.  At the moment, we are only concerned
with the explosion energy, so we use the results in that figure to estimate the
marginalized posterior distributions for explosion energy.  We model
their distributions as a log normal; Table~\ref{tab:eobstable}
provides the mode and uncertainties for the explosion energies.

\begin{table}
\caption{Explosion energies inferred from SN~IIP observations.  $\epsilon$ is the mode, and $\sigma_{\epsilon}$ is the width of the posterior distributions from Figure~2 of \citet{pejcha2015} and Table~4 of \citet{muellerprieto2017}.}
\label{tab:eobstable}
\begin{tabular}{lcc}
\hline
Name & $\epsilon = \log_{10}(E_{\rm obs}/10^{51})$ & $\sigma_{\epsilon}$  \\ 
\hline
SN 2001dc & -1.13 & 0.33\\ 
SN 2013am & -0.98 & 0.25\\ 
SN 1980K & -0.77 & 0.27\\ 
SN 1995ad & -0.62 & 0.23\\ 
SN 2005cs & -0.55 & 0.21\\ 
SN 2009js & -0.51 & 0.43\\ 
SN 2009dd & -0.29 & 0.46\\ 
SN 2012A & -0.28 & 0.10\\ 
SN 2009N & -0.24 & 0.15\\ 
SN 2009bw & -0.20 & 0.13\\ 
SN 2004A & -0.13 & 0.28\\ 
SN 2008in & -0.13 & 0.60\\ 
SN 2004dj & -0.14 & 0.23\\ 
SN 2004et & -0.02 & 0.16\\ 
SN 1996W & 0.07 & 0.28\\ 
SN 2012aw & 0.08 & 0.16\\ 
SN 1999em & 0.15 & 0.15\\ 
SN 2008bk & 0.48 & 0.53\\ 
SN 1992H & 0.53 & 0.29\\ 
SN 1992ba & -0.09 & 0.21\\ 
SN 2002gw & 0.08 & 0.12\\ 
SN 2003B & -0.37 & 0.23\\ 
SN 2003bn & -0.01 & 0.09\\ 
SN 2003E & 0.09 & 0.14\\ 
SN 2003ef & 0.19 & 0.10\\ 
SN 2003fb & 0.03 & 0.14\\ 
SN 2003hd & 0.05 & 0.08\\ 
SN 2003hn & -0.34 & 0.12\\ 
SN 2003ho & 0.26 & 0.10\\ 
SN 2003T & -0.05 & 0.09\\ 
SN 2009ib & -0.27 & 0.08\\ 
SN 2012ec & -0.10 & 0.06\\ 
SN 2013ab & 0.02 & 0.20\\ 
SN 2014ej & -0.05 & 0.12\\ 
SN 2013fs & -0.29 & 0.08\\ 
SN 2014G & -0.17 & 0.08\\ 
ASASSN-14gm & 0.15 & 0.11\\ 
ASASSN-14ha & -0.51 & 0.16\\ 
\hline 
\end{tabular}
\end{table}

Next, we infer the distribution of observed explosion energies by
modeling the mean ($\mu_{\rm obs}$) and width ($\sigma_{\rm obs}$) of the observations.  For a rough estimate,
one may calculate the mean and variance of the modes (2nd column in
Table~\ref{tab:eobstable}).  However, when the uncertainties in the
observations are large, these estimates can easily be biased.  In
particular, the observed variance in the distribution is a convolution
of the true width and the large uncertainties, so simply calculating
the variance of the observations will lead to an over estimation of
the width.  Therefore, we use Bayesian inference to infer the
distribution of explosion energies.  

\begin{figure}
%	% Allowable file formats are eps or ps if compiling using latex
%	% or pdf, png, jpg if compiling using pdflatex
	%\includegraphics[width=\columnwidth]{data/InferObsDist_Posterior.pdf}
	\includegraphics[width=\columnwidth]{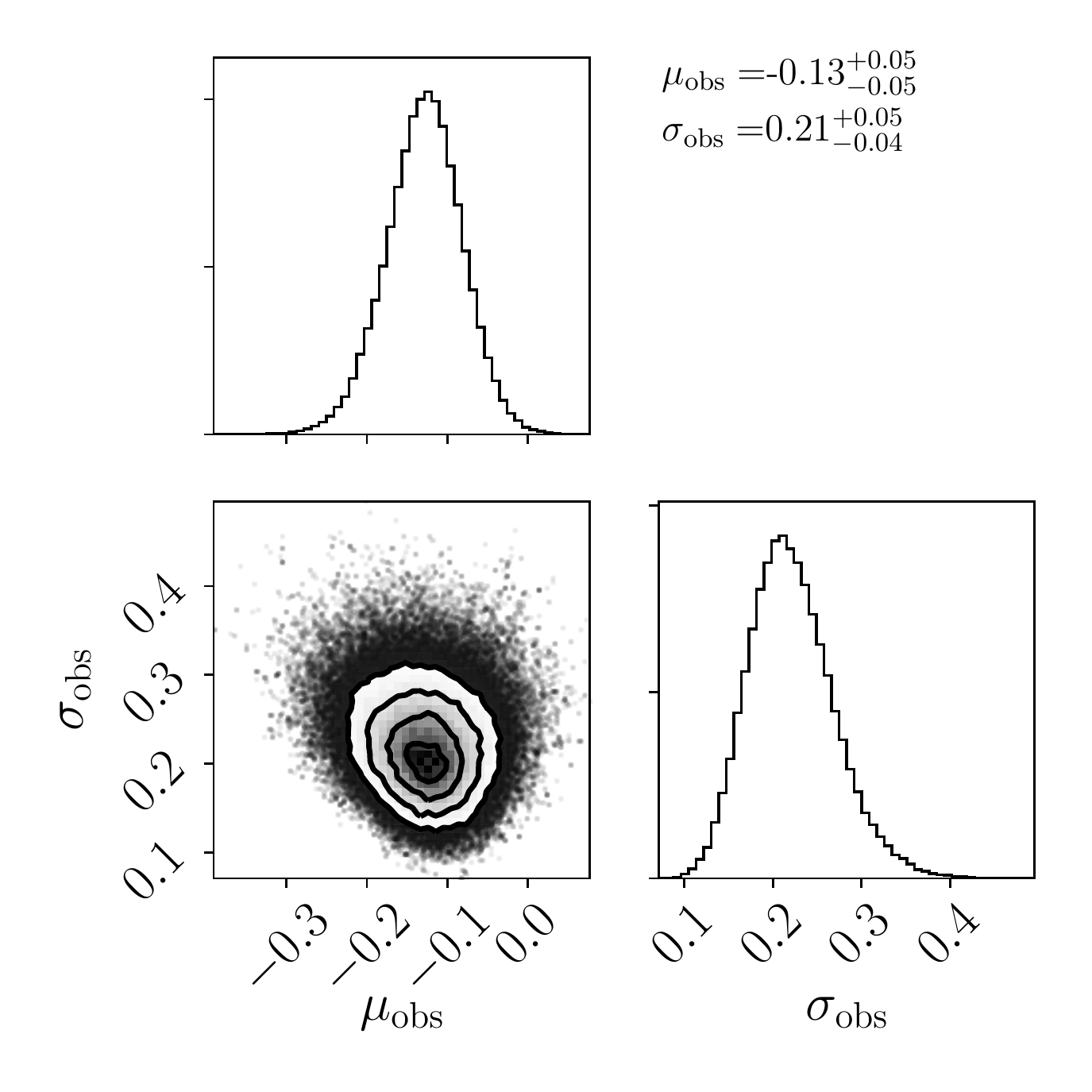}
    \caption{The posterior distribution for $\mu_{\rm obs}$ and
	  $\sigma_{\rm obs}$ in $\log_{10}(E/10^{51}\, {\rm erg})$.  We assume that the SN IIP explosion
	  energies are drawn from a log normal distribution and infer the
	  mean and width.  The mean corresponds to an energy of $\sim 8
	  \times 10^{50}$ erg.}
    \label{fig:EobsPosterior}
\end{figure}

The posterior distribution is
\begin{multline}
\label{eq:obsposterior}
P(\mu_{\rm obs}, \sigma_{\rm obs} | \{ \epsilon_i,\sigma_{\epsilon}
\}) \\ \propto \prod_i
\mathcal{L}(\epsilon_i|\sigma_{\epsilon,i},\mu_{\rm obs},\sigma_{\rm
  obs}) P(\mu_{\rm obs}) P(\sigma_{\rm obs}) \, ,
\end{multline}
where $P(\mu_{\rm obs})$ and $P(\sigma_{\rm obs})$ are uniform
priors.  The likelihood for each observation is
\begin{multline}
\mathcal{L}(\epsilon_i| \sigma_{\epsilon,i},\mu_{\rm obs},\sigma_{\rm
  obs}) \\ = 
\frac{1}{\sqrt{2 \pi (\sigma_{\epsilon,i}^2 + \sigma_{\rm obs}^2)^2}}
e^{-[\epsilon_i - \mu_{\rm
	  obs}]^2/[2(\sigma_{\epsilon,i}^2+\sigma_{\rm obs}^2)^2]} \, .
\end{multline}
To infer this posterior distribution, we use the Markov Chain Monte
Carlo package {\it emcee} \citep{foreman-mackey13}; Figure~\ref{fig:EobsPosterior}
shows the posterior distribution for the model parameters; the mode
and the 68\% highest density intervals (HDI) are $\mu_{\rm obs} = \muobs$ and $\sigma_{\rm obs} = \sigmaobs$.  The
mean corresponds to an energy of $\sim 8 \times 10^{50}$ erg.

\section{Inferred Explosion Energies from CCSN Simulations}
\label{sec:simulations}

To simulate the core-collapse problem with some fidelity, CCSN
simulations must include the following physics: multi-dimensional
hydrodynamics, general relativity, dense nuclear equations of state (EOS),
weak interactions, nuclear reactions, and neutrino transport.  Codes that simulate all of these physics with any fidelity are computationally
expensive.  For example, current three-dimensional neutrino radiation
hydrodynamics simulations require 10s of millions of CPU-hours; on
10,000 cores or more this requires months of computational time for
just one run.  As a result most codes make some approximations.  Even
the most advanced codes require some approximations.  For the purpose of
this study, we only select simulations that meet the following minimum
approximations.  Gravity should include at least a pseudo GR spherical
potential.  The neutrino transport should be a self-consistent
approximation of the Boltzmann equation.  Two examples of such
neutrino transport are 1) solving the Boltzmann equation using discrete
methods along rays, and 2) solving moment equations.  The transport
also should be multi-species, multi-group, and
multi-angle in its approximation.  The individual simulations should
also show signs of approaching a final explosion energy.

The codes that satisfy these technical requirements are CHIMERA,
CoCoNuT-FMT, FORNAX, PROMETHEUS-VERTEX, Zelmani, FLASH, and
the \citet{kuroda2016} code.  However, not
all of these have simulations that explode and asymptotically approach
a final explosion energy.  Only simulations using CHIMERA, FORNAX, and CoCoNuT-FMT satisfy all conditions.  The following is a
brief description of each code, including references that include the
simulation sets.

\subsection{Codes and Simulation Sets Included in This Study}

{\bf CHIMERA:} The full code architecture
and capabilities are presented in \citet{bruenn2006,messer2007,messer2008,bruenn2009,bruenn2013}.  The hydrodynamics solver is a dimensionally-split,
Lagrangian-plus-remap Newtonian scheme with piece-wise parabolic reconstruction. Self-gravity is computed by a multi-pole expansion and the
neutrino transport is computed using ray-by-ray, with multi-group flux
limited diffusion (MGFLD) as the transport solver.   The simulations
of \citet{bruenn2016} use the $K=220$ MeV incompressibility version of the
\citet{lattimer1991} EOS for densities $\rho > 10^{11}$ g cm$^{-3}$.

Using two-dimensional simulations, \citet{bruenn2016} report
explosions of the 12, 15, 20, and 25 M$_{\odot}$
progenitors of \citep{woosley2007}. \citet{lentz2015} report the explosion of the 15
M$_{\odot}$ progenitor in three-dimensional simulations. With only one
three-dimensional simulation, it is difficult to explore the
trends and systematics with mass, etc. Furthermore, the simulation
ends after 440 ms past bounce and 140 ms past the initiation of
positive diagnostic explosion energies.  The explosion energy does not start to plateau,
and thus our extrapolation model for late times would be invalid in
this case. Therefore, we restrict the CHIMERA sample to the 2D simulations of \citet{bruenn2016}.   %For the one progenitor that was simulated in both 2D and 3D XXX.

{\bf CoCoNuT-FMT:} The primary description of this code's architecture
is in \citet{mueller2015}.  The latest advancements for this code are in
\citet{mueller2019}. The hydrodynamics solver for CoCoNuT-FMT solves
the general relativistic hydrodynamics in spherical coordinates on a
unsplit finite-volume mesh.  Fluxes are calculated using an HLLC
Riemann solver, and the metric equations are solved in the extended
conformal flatness approximation with a spherically symmetric metric.
The neutrino transport is multi-group and uses a variable Eddington factor closure and
solves the transport using ray-by-ray.  The transport includes
gravitational redshift but neglects both
velocity dependent terms and inelastic scattering.  However, there is
a Doppler correction to the absorption opacity.

In a three-dimensional simulation, \citet{mueller2015} report the
explosion of the 11.2-M$_\odot$ progenitor of \citet{woosley2002}; for
this simulation, \citet{mueller2015} employ the $K=220$ MeV version of
the \citet{lattimer1991} nuclear EOS. Using Coconut-FMT and a 3D 18-M$_{\odot}$
initial progenitor to provide perturbations, \citet{mueller2017}
produced a perturbation-aided explosion. Most recently,
\citet{mueller2019} produced several explosions of progenitors with
zero-age main sequence (ZAMS) masses between 9.6 M$_\odot$ and 12.5
M$_\odot$. Since the \citet{mueller2019} study
probes a sufficiently high resolution of the mass space, we choose the
diagnostic energies from this set of simulations for our examination.
Since there is only one model that explores perturbation-aided explosions,
we do not include the results of 18-M$_{\odot}$ simulation
\citep{mueller2017} in the final comparison with observations.
However, in section~\ref{sec:Conclusion}, we do discuss the possible
implications of perturbations on explosion energies in simulations. 

{\bf FORNAX:}  The technical details and capabilities of FORNAX are presented in
\citet{skinner2018}.  In summary, this code solves both hydrodynamics
and radiation transport using explicit, finite volume Godunov
schemes.  For gravity, they use a multi-pole solver and
replace the monopole part of the potential with a post-Newtonian
approximation for GR. The transport algorithm is a multi-group, two-moment closure
scheme and uses the M1 moment closure for the Eddington tensor.  Both
the hydro and the transport components calculate the fluxes between cells using approximate Riemann
solvers.  Because the transport is explicit, the time step is limited
by the speed of light across the zone.  In general, the speed of sound
in the protoneutron star is $\sim$1/3 the speed of light, so calculating the
neutrino transport explicitly only increases the number of time steps
by a factor of $\sim$3.  The reductions in calculations for an
explicit transport solver versus an implicit solver more than
compensate for this increase in speed. In general, simulations
involving FORNAX use either the $K=220$ MeV version of the
\citet{lattimer1991} EOS or the SFHo EOS \citep{steiner2013} dense
nuclear equations of state.

There are four primary publications that report CCSN explosions in
FORNAX simulations.  We divide them into two sets, two-dimensional
simulations, FORNAX2D, and three-dimensional simulations, FORNAX3D.  \citet{radice2017} explored explodability of
two-dimensional simulations for the following progenitors: n8.8, u8.1,
z9.6, 9.0, 10.0, 11.0.  The numbers in these models represent the
zero-age main sequence mass.  All models use the K=220 MeV version of
the \citet{lattimer1991} EOS; they also explode in both
one-dimensional and two-dimensional simulations.
\citet{vartanyan2018a} simulated collapse of the 12, 13, 15, 16, 17,
19, 20, 21, and 25 M$_{\odot}$ progenitors \citep{woosley2007}.  For
these simulations, they use the SFHo EOS \citep{steiner2013}.
They reported explosions for the 16, 17, 19, and 20 M$_{\odot}$
progenitors, but only the 16, 17, and 19 M$_{\odot}$ progenitors
provide diagnostic explosion energies that are greater than zero and
approach an asymptotic value.  \citet{vartanyan2019} simulated
three-dimensional
collapse and explosion of the 16 M$_{\odot}$ progenitor.  They find an
explosion, but this simulation has yet to reach positive diagnostic
explosion energies.  More recently, \citet{burrows2019} simulate the
three-dimensional explosions for the same 9, 10, 11, 12, and 13
M$_{\odot}$ progenitors but using the SFHo EOS \citep{steiner2013}.  For the progenitors that are simulated both
in two and three dimensions, the explosion time and diagnostic
explosion energies are very similar.

\subsection{Codes and Simulations Not Included in this Study}

{\bf PROMETHEUS-VERTEX:} \citet{melson2015a} describe the code architecture
for this code.  The hydrodynamics algorithm is a finite volume Godunov
scheme using Riemann solvers to calculate fluxes.
For gravity, the code solves the multi-pole expansion and replaces the
monopole with a pseudo potential that represents a post-Newtonian
approximation to GR.  The neutrino transport solves the Boltzmann
equation on radial rays.

There are two three-dimensional explosions using Prometheus-Vertex.
\citet{melson2015a} report the three-dimensional explosion of a 9.6
M$_{\odot}$ star.  In the same year, \citet{melson2015b} report the
three dimensional explosion of a 20
M$_{\odot}$ star.  These represent the first self-consistent
three-dimensional explosions.  However, the latter is not what one would consider
a fiducial simulation; it explores strange-quark contributions to the
neutrino-nucleon scattering.  The former does explode and begins to
approach asymptotic values at about 400 ms past bounce or 300 ms past the
initiation of explosion.  The final reported explosion energy is
$0.1 \times 10^{51}$ erg, and the final rate of increase is about
$10^{51}$ erg s$^{-1}$.  \citet{summa2016} simulated the explosion of
18 progenitors in two-dimensional simulations.  However, they only
reported diagnostic explosion energies for 4 of the progenitors, and
these did not reach asymptotic values in the explosion energy.  With
only one simulation reaching the final phase of the explosion, it is
difficult to make any systematic conclusions about the performance of
Prometheus-Vertex simulations. Therefore, we do not include Prometheus-Vortex results at this time.
%In the case of PROMETHEUS-VERTEX, only two progenitors were looked at (9.6 M$_\odot$ \citep{melson2015a} and 20 M$_\odot$ \citep{melson2015b} 

{\bf FLASH:}  \citet{oconnor2018} include approximate GR in FLASH, a finite-volume
hydrodynamics code.  The gravity algorithm solves the Newtonian
Poisson's equation via a multi-pole solver and replaces the monopole
term with a post-Newtonian pseudo GR potential.  The
neutrino transport solves the two-moment equations and uses the M1
closure.  The transport is also multi-group, includes
velocity dependence and inelastic scattering.

Using FLASH, \citet{oconnor2018} simulated the collapse of the 12, 15,
20, and 25
M$_{\odot}$ progenitors \citep{woosley2007}.  The 15, 20, and 25
M$_{\odot}$ runs exploded, reaching explosion energies ranging from
0.15 to $0.25 \times 10^{51}$ erg.  However, none reach the
plateau phase in diagnostic explosion energy, so we are not able to
include these results in our comparison.

{\bf Zelmani:} \citet{roberts2016} present the code
architecture. Zelmani is a three-dimensional GR, multi-group
radiation-hydrodynamics code.  The neutrino transport solves the
two-moment equations and uses an M1 closure.  Zelmani also neglects velocity dependence
and inelastic scattering processes.  \citet{ott2018} simulated the
explosions of 12-,
15-, 20-, 27-, and 40-M$_{\odot}$ progenitor models of
\citet{woosley2007} and with the SFHo \citep{steiner2013}
EOS. Though the fidelity of this code meets the
requirements of our analysis, the simulations terminated very shortly
after explosion, and so the explosion curves are far from their
plateau phase.  Thus, we are not able to include these simulations in
this study.

{\bf Kuroda (2016):}  The code presented in \citet{kuroda2016} meets
some of the technical requirements.  The neutrino transport is
multi-group and is a
two-moment scheme using the M1 closure.  The code solves the GR
field equations.  They
simulated collapse of a 15 M$_{\odot}$ progenitor.  However, this
simulation did not explode.

%{\bf ZEUS-MP:}  \citet{takiwaki12} adapted the the ZEUS-MP
%hydrodynamics code to simulte CCSN simulations.  The hydrodynamics
%algorithm is a finite difference scheme and evolves the hydrodynamics
%equations on a staggered mesh.  To resolve shocks, it uses an
%artifical viscosity scheme.
%The neutrino transport
%is the IDSA \jwm{If this is IDSA,
%  then I'm not going to include it.}  \citet{takiwaki12} had a
%successful explosion with an 11.2 M$_\odot$ model but did not report
%explosion energies. 

%{\bf Takiwaki Code (ZUES-MP?) (2012):}
%\begin{itemize}
%\item MHD algorithm(MOCCT)
%\item variable tensor Eddington factor
%\item 3D
%\item IDSA neutrino transport (ignores velocity-dependence, GR, and inelasticity, stitches together the opaque and transparent realms)
%\item Exploded an 11.2 SM model
%\item did not report explosion energies 

%\end{itemize}

\section{Extrapolating Simulated Explosion Energies to Late Times}
\label{sec:extrapolating}

The explosion energies in many of the multi-dimensional simulations
are still increasing when the simulations terminate.  At the same time,
most appear to be approaching an asymptotic value.  Therefore, at best,
the reported diagnostic explosion energies are a minimum.  Here, we
note that the energy evolution for all simulations follows a common
functional form.  We suggest a simple model for explosions driven
by neutrino heating.  This model leads to a simple functional form
that is a good fit to the simulations.  Here we use this functional
form to extrapolate the explosion energy to infinite time, $E_{\infty}$.

\begin{figure}
%	% Allowable file formats are eps or ps if compiling using latex
%	% or pdf, png, jpg if compiling using pdflatex
	%\includegraphics[width=\columnwidth]{data/ExplosionEnergyvsTime.pdf}
	\includegraphics[width=\columnwidth]{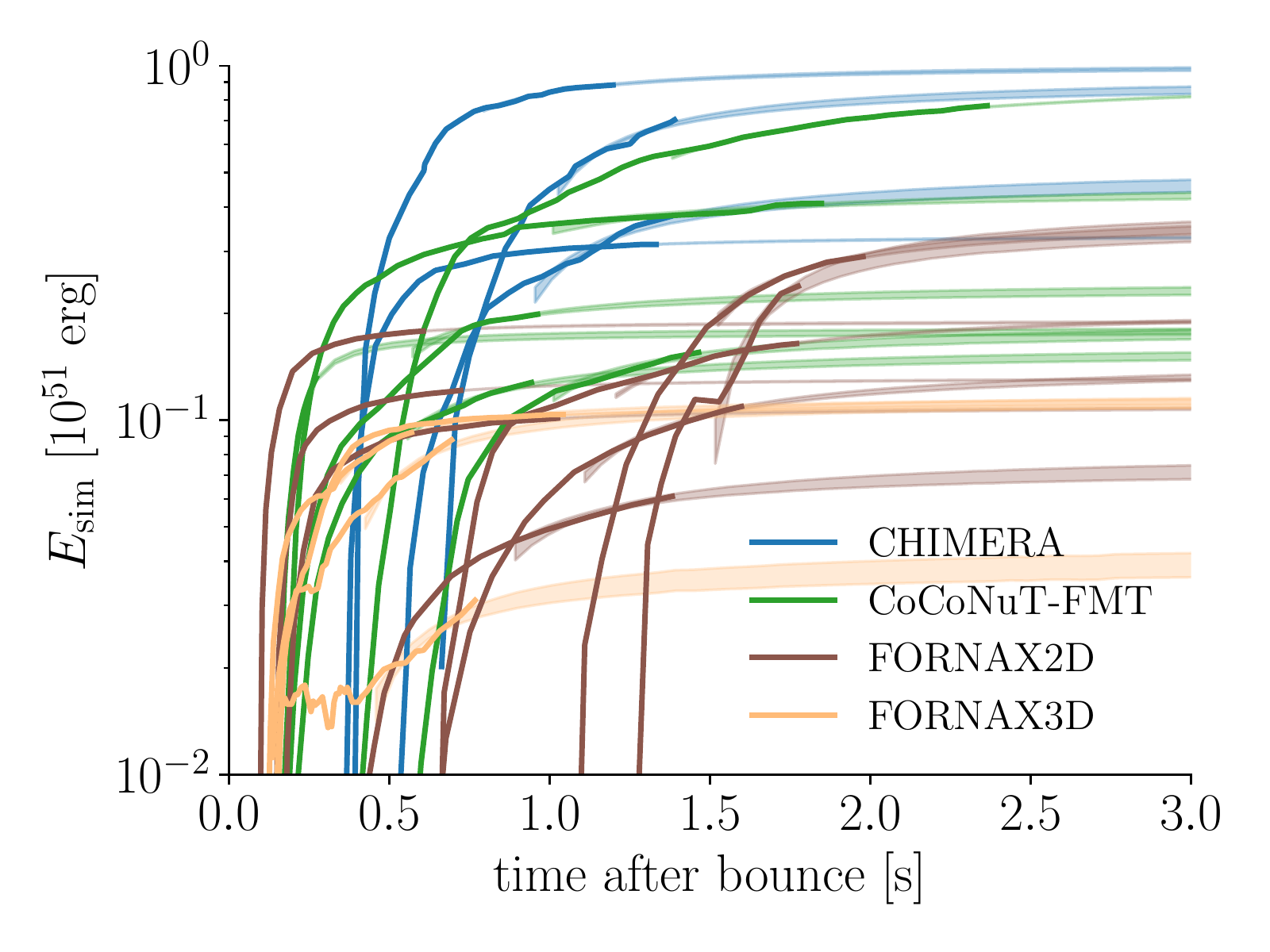}
    \caption{Simulated explosion energies ($E_{\rm sim}$) vs. time after bounce.  The thick solid
	  lines represent the diagnostic explosion energies for CHIMERA
	  (blue),
	  CoCoNuT-FMT (green), FORNAX2D (brown), and FORNAX3D (yellow).
	  Assuming that the growth of explosion energy is dominated by
	  neutrino power, we propose a simple extrapolation of the
	  explosion energy curve.  See eq.~(\ref{eq:eexpvst}).  The wide bands
	  represent a 68\% confidence interval extrapolation.  From this
	  extrapolation we infer an explosion energy after infinite time,
	  $E_{\infty}$.  The model appears to be a good fit for nearly all
	simulations except the very under-energetic model in the FORNAX3D
	set.  That simulation corresponds to the 10.0 M$_{\odot}$ progenitor
	and is likely still developing the explosion profile even after
	750 ms.}
    \label{fig:Esimvst}
\end{figure}

Figure~\ref{fig:Esimvst} shows the diagnostic explosion energies in
units of $10^{51}$ erg as a function of time after bounce.  The solid
lines represent the results of multi-D simulations, and the bands at
the end of each curve represents our extrapolation.  Blue represents
CHIMERA simulations, green represents CoCoNuT-FMT, brown
represents FORNAX2D, and yellow represents FORNAX3D.  All curves rise
and show signs of asymptotically approaching a finite explosion energy,
$E_{\infty}$.  In a preliminary analysis, we considered two functional
forms: an exponential and a power-law.  These crude initial
comparisons suggests that the energy curves asymptote via a power-law
and not an exponential.  Using this crude analysis as a guide, we now
suggest a model for the explosion energy curve and derive the
functional form.

If neutrinos are primarily driving the explosion, then one might
expect the rate of growth of explosion energy to be roughly proportional to the neutrino power.
\begin{equation}
\label{eq:dedt}
\frac{d E_{\rm exp}}{dt} \approx L_{\nu} \tau \, ,
\end{equation}
where $L_{\nu}$ is the neutrino luminosity and $\tau = \int \rho
\kappa \, dr$ is the optical depth to neutrino absorption in the
region of net neutrino heating.  $\kappa$ is the neutrino absorption
cross section per unit mass, $\kappa \approx \sigma/m_p$.  

A few straightforward assumptions lead to a simple function for $E_{\rm sim}
(t)$.  First, during the explosion, we assume that the optical depth
is roughly $\tau \sim \kappa M_{\rm gain}/R_s^2$, where $M_{\rm gain}$
is the mass in the gain region and and $R_s$ is the shock radius.
Making the simplest assumptions, we assume that $L_{\nu}$ and $M_{\rm
  gain}$ are roughly constant during the last stage of explosion
development.  In addition, we assume that $R_s = v_s t$, and that the
shock velocity, $v_s$ is also constant.  Integrating
eq.~(\ref{eq:dedt}) leads to the following functional form
\begin{equation}
\label{eq:eexpvst}
E_{\rm exp}(t) = E_{\infty} - \frac{A}{t} \, ,
\end{equation}
Formally, $A$ is proportional to $L_{\nu} \kappa M_{\rm gain}/v_s^2$,
but we do not have access to these values for all of the simulations.
Therefore, in our extrapolations, we fit only for two parameters,
$E_{\infty}$ and $A$.

We use Bayesian inference to find the best fit values for $E_{\infty}$
and $A$.  The posterior distribution for $E_{\infty}$ and $A$ is
\begin{equation}
\label{eq:einfposterior}
P(E_{\infty},A,\sigma | \left \{ E_{{\rm sim},i} \right \} ) \propto
\mathcal{L}(E_{\rm sim,i} | E_{\infty}, A, \sigma)
P(E_{\infty}) P(A) P(\sigma) \, .
\end{equation}
It is unclear what the variance $\sigma^2$ is for the simulations.  Therefore, we
include $\sigma$ as an unknown nuisance parameter and simply
marginalize over all possible values to infer the posterior
distribution for $E_{\infty}$ and $A$, $P(E_{\infty},A)$.
With little prior information about any of these parameters, we choose
uniform priors for $P(E_{\infty})$, $P(A)$, and $P(\sigma)$.  To model the
likelihood, we assume a Gaussian distribution for each simulation data
point:
\begin{equation}
\mathcal{L}(\left \{ E_{{\rm sim},i}\right \}|E_{\infty,A}) = \prod_{i}
\mathcal{N}_i(E_{{\rm sim},i}|E_{\rm exp}(t_i,E_{\infty},A),\sigma) \, ,
\end{equation}
where
\begin{multline}
\mathcal{N}_i(E_{{\rm sim},i}|E_{\rm exp}(t_i,E_{\infty},A),\sigma) = \\
\frac{1}{\sqrt{2 \pi} \sigma} e^{- [ E_{{\rm sim},i} - E_{\rm
	exp}(t_i,E_{\infty},A) ]^2/ [ 2 \sigma^2 ] } \, .
\end{multline}
The mean is the modeled explosion energy, $E_{\rm exp}(t)$
eq.~(\ref{eq:eexpvst}).  The unknowns to infer are the asymptotic
explosion energy, $E_{\infty}$, the parameter for the $1/t$ term, $A$,
and the unknown variation within each simulation, $\sigma$.

When inferring these parameters, we only fit the last half of the
energy curve.  The primary assumptions of the evolution model assume that $L_{\nu}$, $M_{\rm gain}$, and $v_s$ are
constant.  If these assumptions are appropriate at all, they are likely valid in the
last part of the explosion energy evolution.  To
perform these inferences, we use Markov Chain Monte-Carlo Bayesian
inference package emcee \citep{foreman-mackey13}.
The bands extrapolating the energy curves in Figure~\ref{fig:Esimvst}
show the resulting inferences.  The width of the band represents the
68\% highest density confidence interval (HDI) for these fits.

Table~\ref{tab:explosionenergies} summarizes the set of simulation
explosion energies.  The first row gives the progenitor as presented
in the simulation papers.  Each progenitor name conveniently indicates
the ZAMS mass in M$_{\odot}$.  The second column reports the final
explosion energy of the simulation, $E_{\rm sim}(t_{\rm end})$.  The
third column presents the end of the simulation in seconds
after bounce, $t_{\rm end}$.  Finally, column four shows the mode of the
extrapolated explosion energy, $E_{\infty}$.

\begin{table}
\caption{Explosion energies for 2D and 3D CCSN simulations. The simulations for CHIMERA are 2D, CoCoNut-FMT are 3D, FORNAX2D are 2D, and FORNAX3D are 3D.  The 18ProgConv model represents the explosion of the 18 M$_{\odot}$ progenitor that includes pre-collapse perturbations due to O-shell burning \citep{mueller2017}.  Since the initial conditions are different from the other CoCoNuT-FMT simulations, we do not include 18ProgConv in the rest of the explosion energy analysis.  However, we do discuss the possible ramifications of progenitor convection in section \ref{sec:Conclusion}. See text for references and further discussion.}
\label{tab:explosionenergies}
\begin{tabular}{lccc}
\hline
Progenitor & $E_{\rm sim}(t_{\rm end})$ [$10^{51}$ erg] & $t_{\rm end}$ [s] & $E_{\infty}$ [$10^{51}$ erg] \\ 
\hline
\multicolumn{4}{c}{CHIMERA} \\ 
\hline 
12 & 0.31 & 0.97 & 0.34\\ 
15 & 0.88 & 0.81 & 1.03\\ 
20 & 0.38 & 0.84 & 0.50\\ 
25 & 0.70 & 0.73 & 0.93\\ 
\hline 
\multicolumn{4}{c}{CoCoNuT-FMT} \\ 
\hline 
11.2 & 0.13 & 0.77 & 0.16\\ 
s11.8 & 0.20 & 0.78 & 0.24\\ 
s12.5 & 0.16 & 0.90 & 0.19\\ 
z12 & 0.41 & 1.68 & 0.47\\ 
z9.6 & 0.13 & 0.12 & 0.18\\ 
18ProgConv & 0.77 & 1.96 & 0.98\\ 
\hline 
\multicolumn{4}{c}{FORNAX2D} \\ 
\hline 
11.0 & 0.11 & 0.98 & 0.15\\ 
9.0 & 0.06 & 0.98 & 0.08\\ 
n8.8 & 0.18 & 0.51 & 0.19\\ 
u8.1 & 0.10 & 0.85 & 0.11\\ 
z9.6 & 0.12 & 0.58 & 0.13\\ 
16 & 0.16 & 1.13 & 0.21\\ 
17 & 0.29 & 0.90 & 0.39\\ 
19 & 0.24 & 0.52 & 0.39\\ 
\hline 
\multicolumn{4}{c}{FORNAX3D} \\ 
\hline 
9.0 & 0.10 & 0.91 & 0.11\\ 
10.0 & 0.03 & 0.62 & 0.04\\ 
11.0 & 0.09 & 0.44 & 0.12\\ 
12.0 & 0.09 & 0.54 & 0.12\\ 
\hline 
\end{tabular}
\end{table}

\begin{figure*}
%	% Allowable file formats are eps or ps if compiling using latex
%	% or pdf, png, jpg if compiling using pdflatex
	%\includegraphics[width=\columnwidth]{data/ObsAndTheoryPDF0.pdf}
	%\includegraphics[width=\columnwidth]{data/ObsAndTheoryPDF0.pdf}
	%\includegraphics[width=\columnwidth]{data/ObsAndTheoryPDF1.pdf}
	%\includegraphics[width=\columnwidth]{data/ObsAndTheoryPDF2.pdf}
	%\includegraphics[width=\columnwidth]{data/ObsAndTheoryPDF3.pdf}
	\includegraphics[width=\columnwidth]{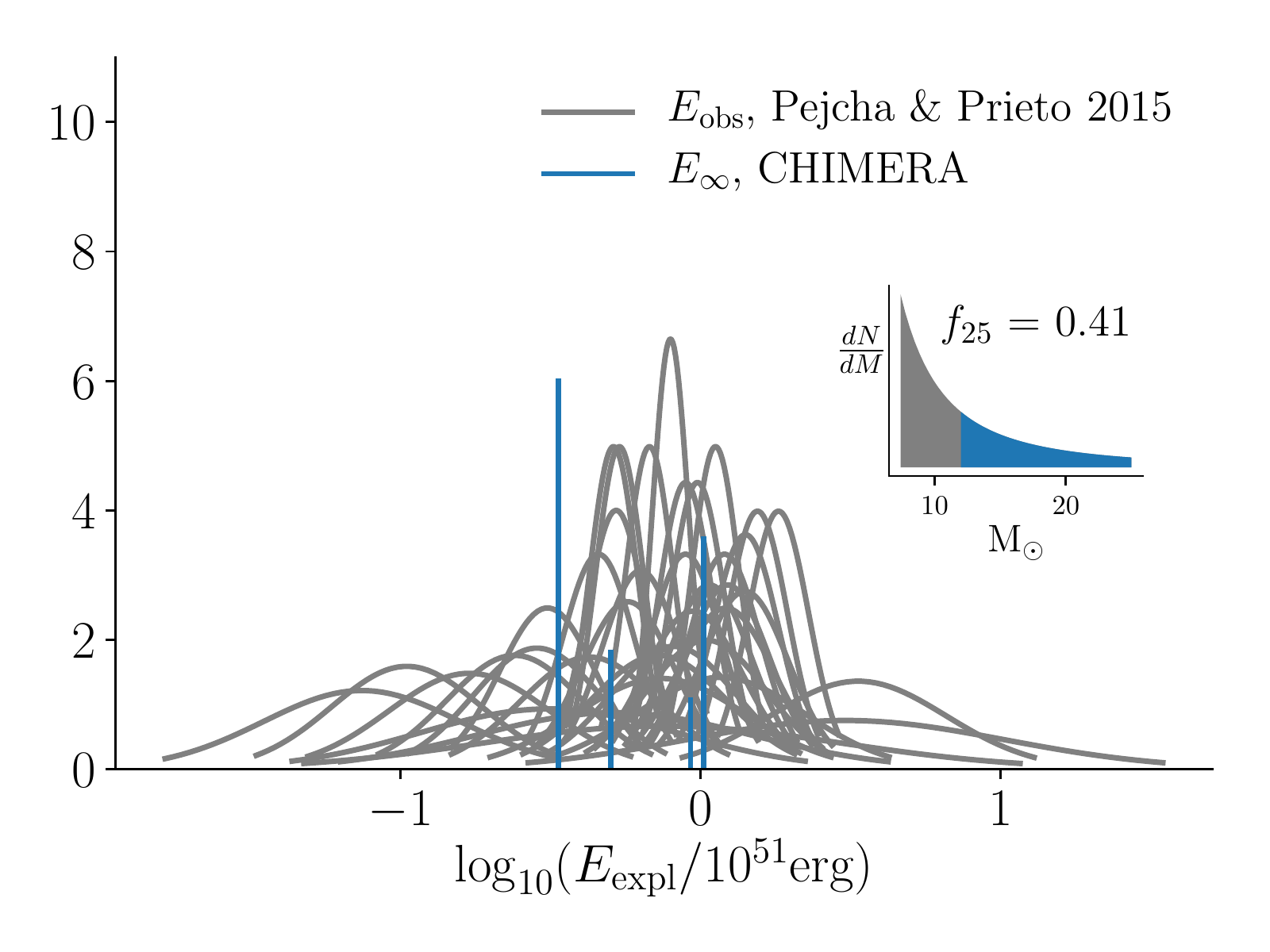}
	\includegraphics[width=\columnwidth]{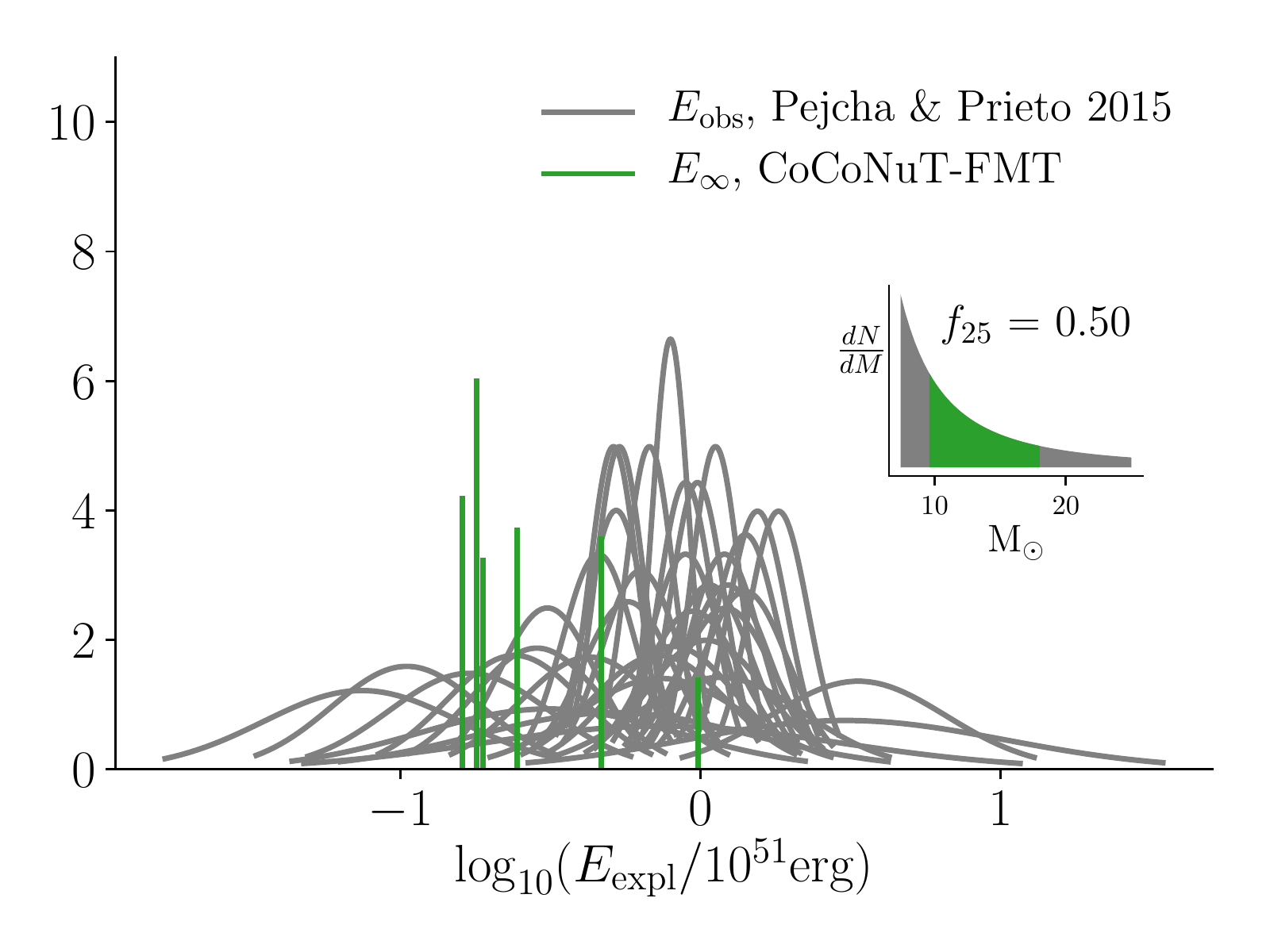}
	\includegraphics[width=\columnwidth]{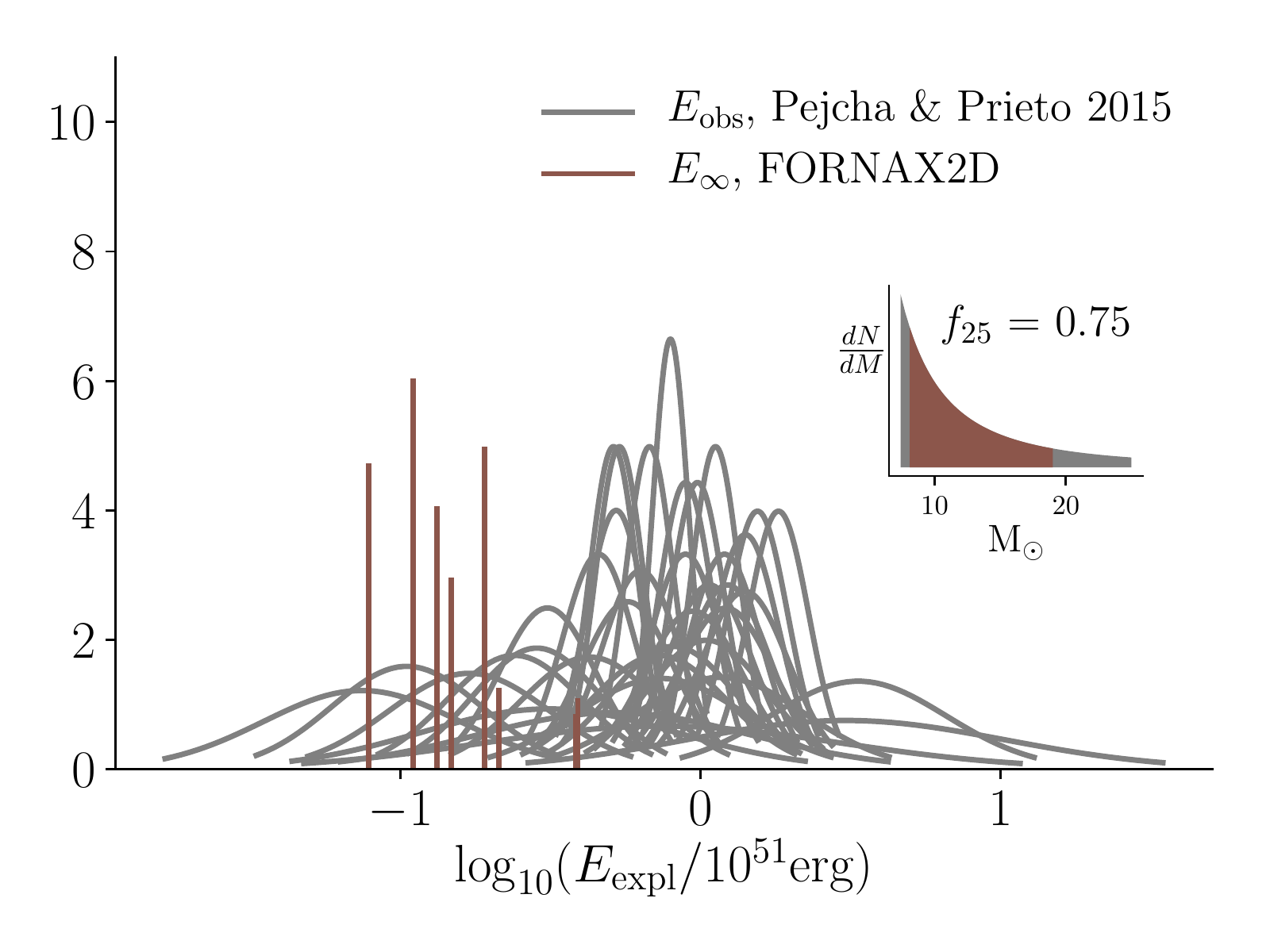}
	\includegraphics[width=\columnwidth]{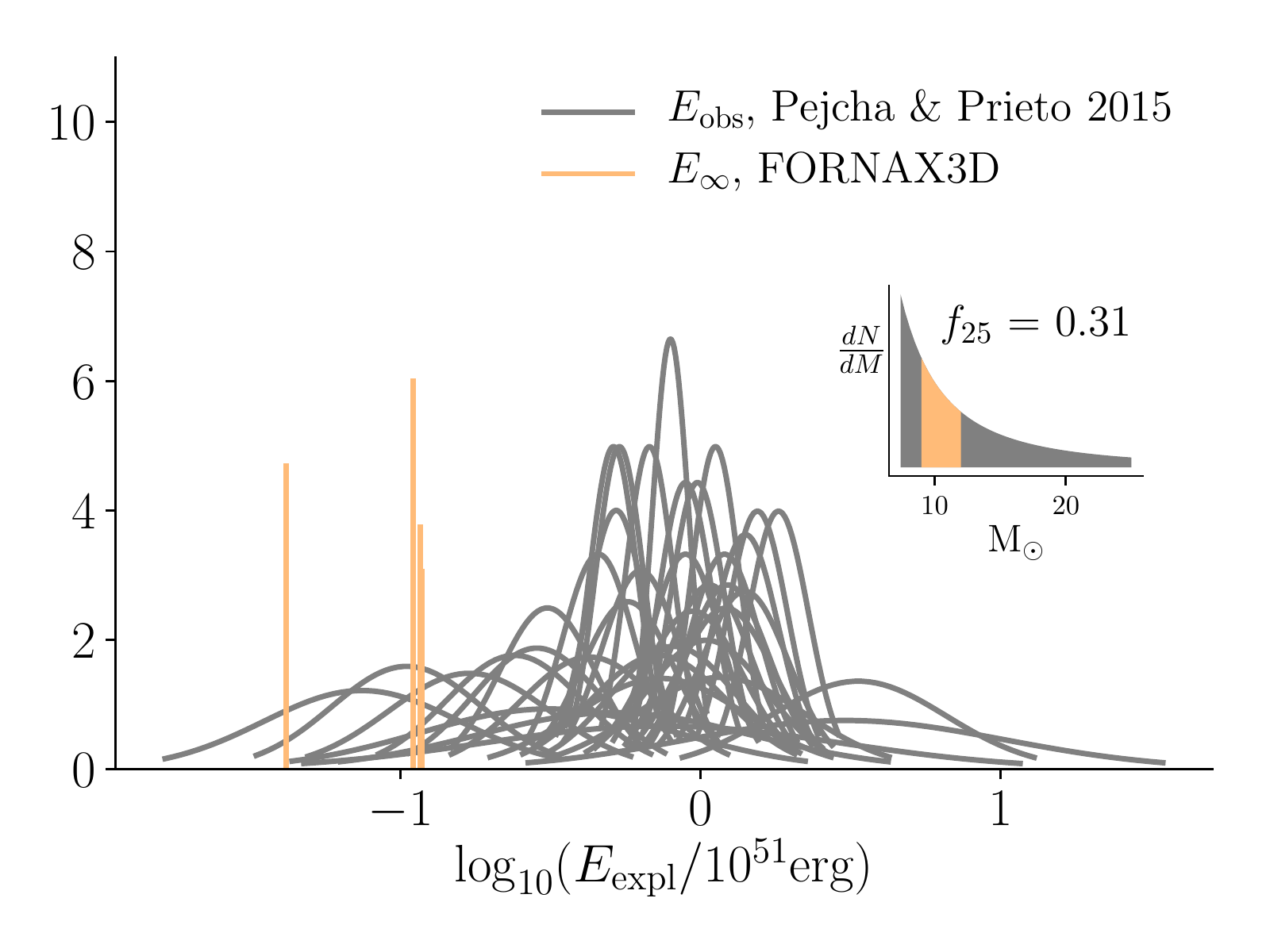}
    \caption{The probability density functions for $E_{\rm obs}$
	  and $E_{\infty}$.  $E_{\rm obs}$ is an estimate for the
	  marginalized observed explosion energy from \citet{pejcha2015}.
	  $E_{\infty}$ represents the extrapolated explosion energies for
	  four simulation sets: CHIMERA represents the two-dimensional
	  simulations of \citet{bruenn2016}; CoCoNuT-FMT represents the
	  three-dimensional simulations of \citet{mueller2015}, \citet{mueller2017}, and
	  \citet{mueller2019}; FORNAX2D represents the two-dimensional
	  simulations of \citet{radice2017} and \citet{vartanyan2018a};
	  FORNAX3D represents the three-dimensional simulations of \citet{burrows2019}.  The
	heights of $E_{\infty}$ represent a weighting due to the IMF.
	The inset and $f_{25}$ represents the fraction of the IMF that the simulations
	have sampled between 7.4 and 25 M$_{\odot}$.  This range roughly
	represents the expected progenitors for SN IIP.}
    \label{fig:EobsvsEsim}
\end{figure*}

The four panels in Figure \ref{fig:EobsvsEsim} compare $E_{\infty}$
with an estimate for the observed explosion energies.  The vertical lines
indicate the estimated simulation explosion energy.  The height of the
lines are proportional to $M^{-2.35}$, representing the initial mass
distribution.  In other words, the height represents the fraction of stars that would explode with that energy
within the simulated set.  The gray lines in each represent the
``marginalized'' inferred explosion energies from observations
\citep{pejcha2015}.  The term marginalized is in quotes because
without the original posterior distributions, we performed a crude
marginalization using Figure~2 from \citet{pejcha2015}.
The fraction $f_{25}$ in each figure represents the fraction of the
IMF that each code has simulated from 7.4 to 25 M$\odot$.  The minimum
corresponds to the minimum mass for CCSNe \citep{diaz-rodriguez2018},
the maximum corresponds to the maximum mass for SN IIP
\citep{smartt2015,davies2018}.  For further discussion on these limits see
section \ref{sec:Conclusion}.

%\begin{figure}
%	% Allowable file formats are eps or ps if compiling using latex
%	% or pdf, png, jpg if compiling using pdflatex
%	\includegraphics[width=\columnwidth]{data/ObsAndTheoryPDF1.png}
%    \caption{The probability density functions for $E_{\rm obs}$
%	  and $E_{\infty}$.  $E_{\infty}$ represent the three-dimensional
%	  simulations of \citet{melson2015} and \citet{mueller2018}.}
%    \label{fig:EobsvsEsim1}
%\end{figure}

%\begin{figure}
%	% Allowable file formats are eps or ps if compiling using latex
%	% or pdf, png, jpg if compiling using pdflatex
%	\includegraphics[width=\columnwidth]{data/ObsAndTheoryPDF2.png}
%    \caption{The probability density functions for $E_{\rm obs}$
%	  and $E_{\infty}$.  $E_{\infty}$ represent the two-dimensional
%	  simulations of \citet{Radice2017} and \citet{vartanyan2018a}.}
%    \label{fig:EobsvsEsim2}
%\end{figure}

Upon first glance, the CHIMERA set appears to be most
consistent with the observations, and the FORNAX sets are the least
consistent.  However, the CHIMERA simulations mostly use the highest
mass progenitors.  Below, we note a correlation between explosion
energy and progenitor mass for the FORNAX2D set.  Therefore, the CHIMERA
results may actually represent the highest explosion energies when a
full range of progenitors are considered.  In other words, the range of
progenitor masses simulated represents a possible bias for each
simulation set.  Below, we model the explosion energy as a function of
progenitor mass to account for this possible bias.

\section{Inferring the Full Simulation Explosion Energy Distribution}
\label{sec:InferEsimDistribution}

\begin{figure*}
%	% Allowable file formats are eps or ps if compiling using latex
%	% or pdf, png, jpg if compiling using pdflatex
	%\includegraphics[width=\textwidth]{data/EsimVsMass.pdf}
	\includegraphics[width=\textwidth]{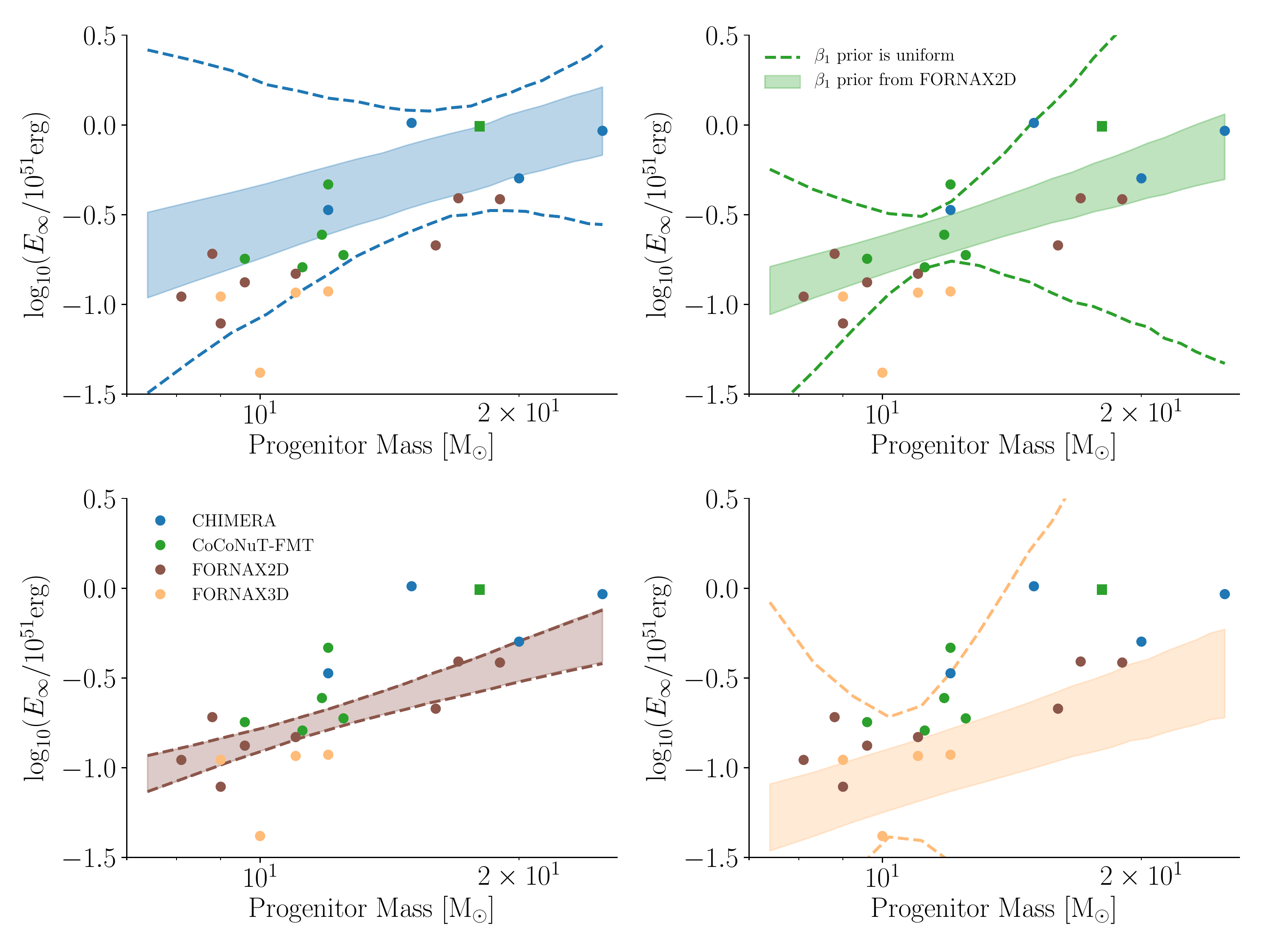}
    \caption{Extrapolated explosion energy for simulations
	  vs. progenitor mass.  CCSN simulations are computationally
	  expensive, and so there are few simulations to compare to
	  observations.  SN IIP are expected to have progenitor masses
	  between 7.4 and 25 M$_{\odot}$.  Therefore, we infer a function relating
	  $E_{\infty}$ to the progenitor mass.  Later, we use this
	  function to infer the distribution of simulated explosion
	  energies between 7.4 and 25 M$_{\odot}$.  We fit a line in log
	  space.  The region between the dashed lines represent the 68\%
	  confidence interval for this slope.  For all sets except
	  FORNAX2D, there are too few simulations to constrain the slope.
	  The solid band represents the 68\% confidence interval when
	  using the slope from the FORNAX2D set as a prior for the other
	  sets. The green square represents the 18ProgConv model of the
	  CoCoNuT-FMT set, and since it is the only model that includes
	  perturbations due to O-shell burning, it is not included in the
	  explosion energy vs. mass fit.}
    \label{fig:Einfvsmass}
\end{figure*}

The simulation sets have not yet sampled the whole range of
progenitors from 7.4 to 25 M$_{\odot}$.  Therefore, the explosion
energies in Table~\ref{tab:explosionenergies} represent a biased sample.  For example,
Figure~\ref{fig:Einfvsmass} plots $E_\infty$ (dots) vs. progenitor mass.  It is apparent
that the CHIMERA set includes mostly high mass progenitors, the
CoCoNuT-FMT set includes mostly the middle, and the FORNAX2D set has
simulated a larger range.  Note that the 18ProgConv model (green
square) is omitted in the analysis for CoCoNuT-FMT.  In this section, we infer a relationship
between explosion energy and progenitor mass.  Then we use this
inference to infer the full distribution of explosion energies between
7.4 and 25 M$_{\odot}$. 

The simplest assumption is that the explosion energy is proportional
to some power of the progenitor mass.  In fact, for SN IIP observations
\citet{pejcha2015} infer that the explosion energy is proportional to
a power of the ejecta mass, $E_{\rm exp} \propto M_{\rm
  ej}^{1.81^{0.45}_{-0.34}}$.  Therefore, the most natural model to
assume for the correlation is
\begin{equation}
\label{eq:energyvsmass}
\log(E_\infty/10^{51} \, {\rm erg}) = \beta_0 + \beta_1 \log(M/10 \, {\rm M}_{\odot})\, .
\end{equation} 
To infer the parameters, $\beta_0$ and $\beta_1$, we use {\it emcee} \citep{foreman-mackey13} to
infer the following posterior distribution
\begin{multline}
\label{eq:posteriormodel}
P(\beta_0,\beta_1,\sigma|\{ E_{\infty,i}\}, \{ M_i\}) \propto \\
\prod_i \mathcal{L}(E_{\infty,i}|M_i,\beta_0,\beta_1,\sigma)
P(\beta_0) P(\beta_1) P(\sigma) \, ,
\end{multline}
where the likelihood for the simulated explosion energy $E_{\infty_i}$
is
\begin{multline}
\mathcal{L}(E_{\infty,i}|M_i,\beta_0,\beta_1,\sigma) = \\
\frac{1}{\sqrt{2 \pi} \sigma} 
e^{-[ \log(E_{\infty,i}/B) - \beta_0 - \beta_1\log(M_i/{\rm M}_{\odot}) ]^2/(2 \sigma)} \, .
\end{multline}
The variation in the simulated energies, $\sigma$, is an unknown
nuisance parameter.  The priors, $P(\beta_0)$, $P(\beta_1)$, and
$P(\sigma)$, are all assumed to be uniform.

\begin{figure}
%	% Allowable file formats are eps or ps if compiling using latex
%	% or pdf, png, jpg if compiling using pdflatex
	%\includegraphics[width=\columnwidth]{data/LinearModel_Posterior2.pdf}
	\includegraphics[width=\columnwidth]{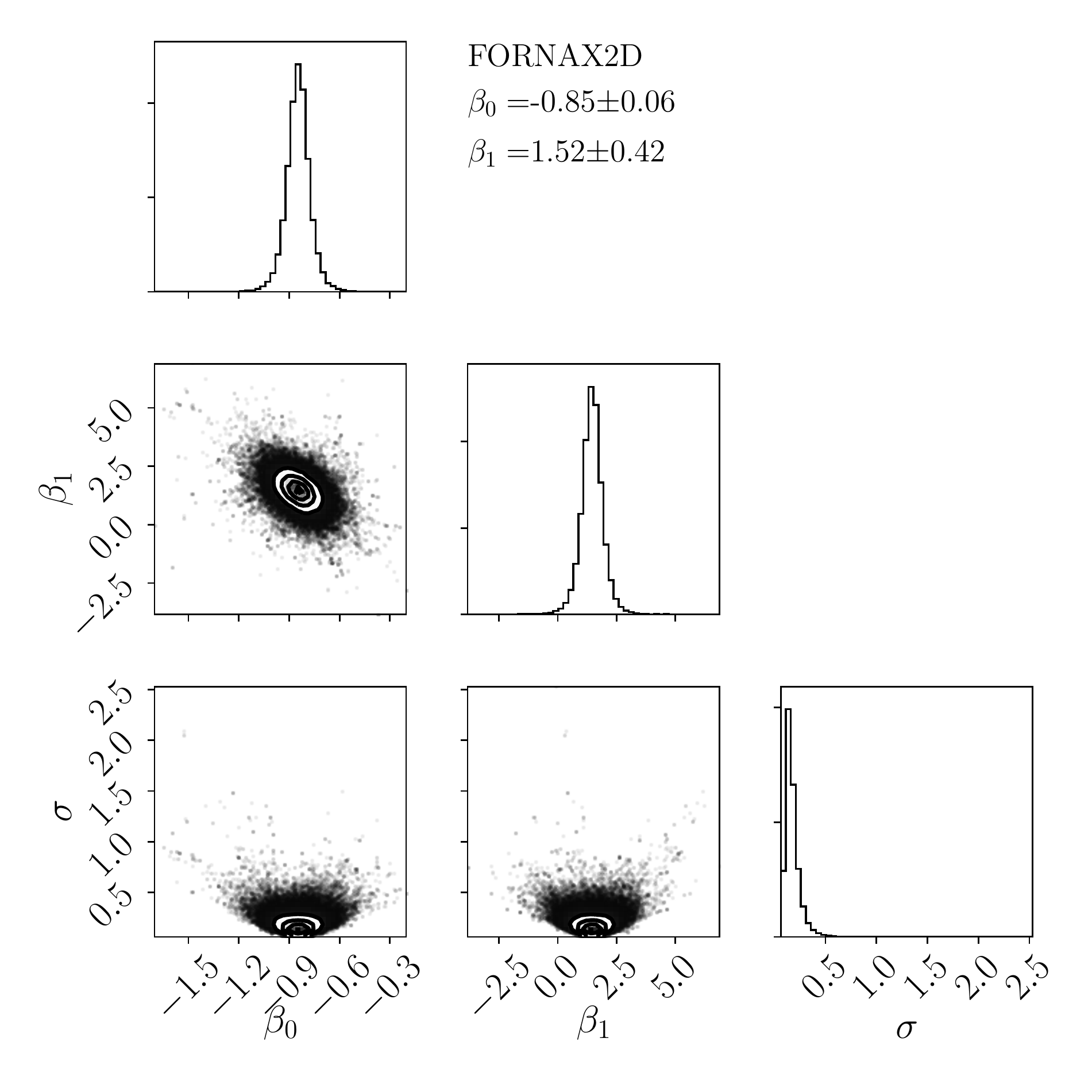}
    \caption{Posterior distribution for the fitting parameters in
	  explosion energy as a function of progenitor mass.  $\beta_0$ is
	  the explosion energy of a 10 M$_{\odot}$ progenitor, and
	  $\beta_1$ is the power-law slope.  See
	  eq.~(\ref{eq:energyvsmass}) for the definition of these
	  parameters.  This posterior distribution is for the FORNAX2D
	  simulation set.}
    \label{fig:Einfvsmassposterior}
\end{figure}

Figure~\ref{fig:Einfvsmassposterior} shows the posterior distribution
for the FORNAX2D simulation set.  The marginalized parameters are $\beta_0 =
\fornaxTwoDbetazero$ and $\beta_1 = \fornaxTwoDbetaone$.  The values are the modes, and the
uncertainties are the 68\% highest density intervals (HDI).  For the
CHIMERA, CoCoNuT-FMT, and FORNAX3D simulation sets, there are far too few
simulations to adequately constrain the slope.  Therefore, we use the
$\beta_1$ distribution for FORNAX2D as the prior for the other two
sets.  The marginalized parameters for CHIMERA are 
$\beta_0 = \chimerabetazero$ and $\beta_1 = \chimerabetaone$; 
the marginalized parameters for CoCoNuT-FMT are 
$\beta_0 = \coconutbetazero$ and $\beta_1 = \coconutbetaone$; 
the marginalized parameters for FORNAX3D are
$\beta_0 = \fornaxThreeDbetazero$ and $\beta_1 =
\fornaxThreeDbetaone$.  The inference for CoCoNuT-FMT does not include
the 18ProgConv model.  The lines in
Figure~\ref{fig:Einfvsmass} represent the distribution of possible
functions.  For each MCMC sample of $\beta_0$ and $\beta_1$, we
calculate $E_{\infty}$ as a function of $M$.  Then we calculate the
68\% HDI for $E_{\infty}$.
Within the confidence intervals, the exponent ranges from linear to quadratic,
$E_{\infty} \propto M^{1-2}$.  

Using observations, \citet{pejcha2015} infer the relationship between the explosion energy
($E_{\rm exp}$ in their manuscript) and the ejecta mass $M_{\rm ej}$.
Since the neutron star that is left behind is only 1.4 M$_{\odot}$,
and there is not much mass loss for the red supergiant progenitors
they consider, the ejecta mass is similar to the ZAMS progenitor
mass.
They find that $E_{\rm exp} \propto M_{\rm
  ej}^{1.81^{+0.45}_{-0.34}}$, which is consistent with our inference.

\begin{figure}
%	% Allowable file formats are eps or ps if compiling using latex
%	% or pdf, png, jpg if compiling using pdflatex
	%\includegraphics[width=\columnwidth]{data/EsimPosterior.pdf}
	\includegraphics[width=\columnwidth]{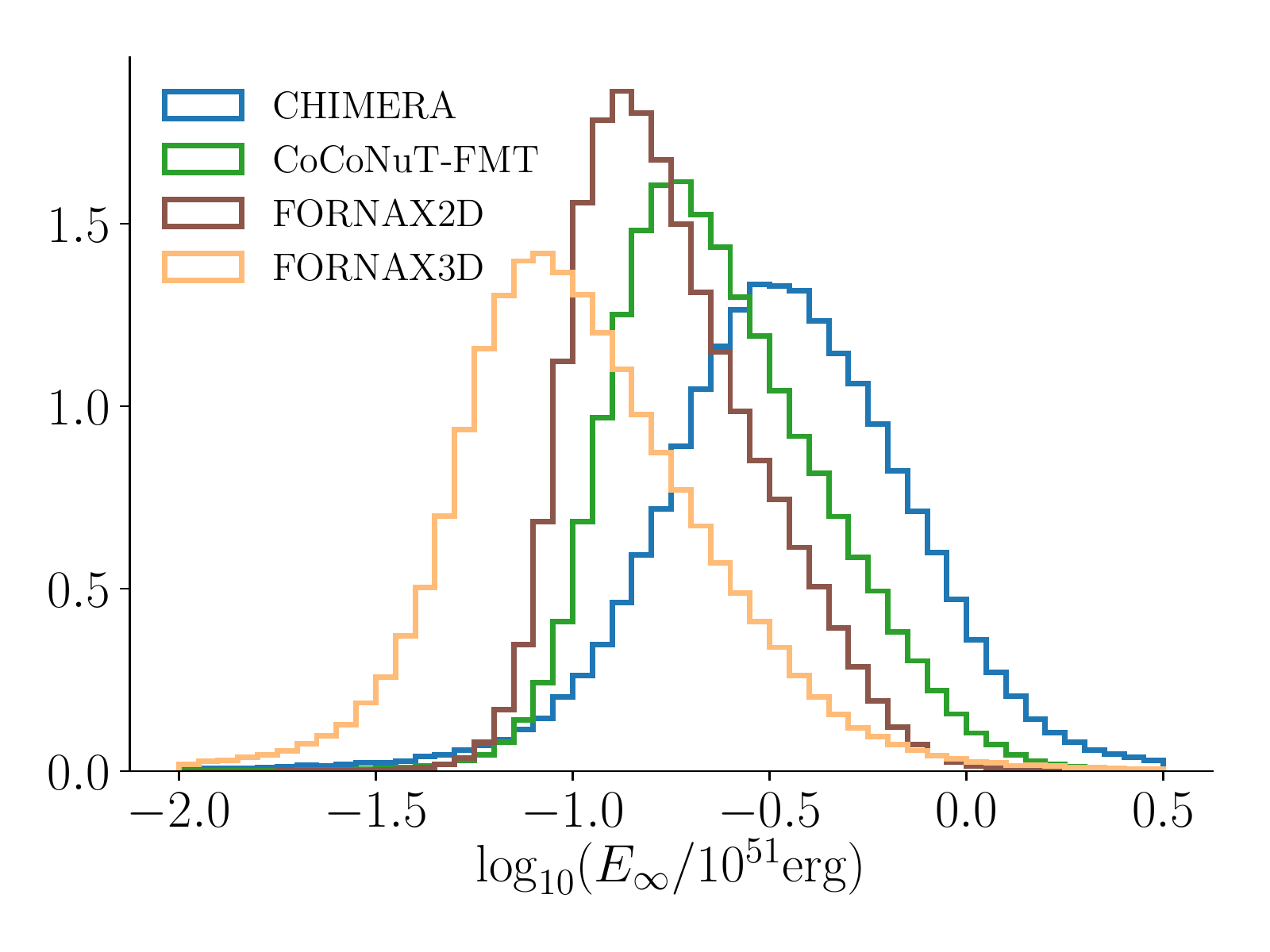}
    \caption{Posterior distribution of simulated explosion energies for each
	  simulation set.  This inference assumes that every progenitor between 7.4
	  and 25 M$_{\odot}$ explodes and that the explosion energy is a function of
	  progenitor mass as given by eq.~(\ref{eq:energyvsmass}) and the fits in
	  Figures~\ref{fig:Einfvsmass} \& \ref{fig:Einfvsmassposterior}.}
    \label{fig:EsimPosterior}
\end{figure}

Next, we extrapolate to infer the explosion energy distribution for
the range of progenitor masses between 7.4 and 25 M$_{\odot}$.   The posterior for the explosion energies is
\begin{multline}
P(E_{\infty}) = \int P(E_{\infty}|\beta_0,\beta_1,\sigma,M) \\ \cdot
P(\beta_0,\beta_1,\sigma) \cdot P(M) \, dM d\beta_0 d\beta_1 \, ,
\end{multline}
where $P(\beta_0,\beta_1,\sigma)$ is the posterior distribution,
eq.~(\ref{eq:posteriormodel}), for the model parameters, and $P(M)$
is the progenitor mass distribution.  For this study, we assume that
$P(M) \propto M^{-2.35}$.  To determine $P(E_{\infty})$, we take a sample
of $(\beta_0,\beta_1,\sigma)$ from MCMC posterior distribution, draw a
mass from the IMF distribution, $P(M) \propto M^{-2.35}$, and evaluate
$E_{\infty}$ using eq.~(\ref{eq:energyvsmass}).  The resulting
distributions for each code are in Figure~\ref{fig:EsimPosterior}.  On
average, all simulations sets exhibit explosion energies that are
significantly lower than the observations.  In the next section, we
quantify the difference between simulations and observations.

%Figure~\ref{fig:EsimPosterior} shows the posterior distribution for
%the inferred 

\section{Comparing Observations with Multidimensional Simulations}
\label{sec:compare}

Since the average explosion energies of the simulations are lower
than the observations, we develop a model to infer the missing
explosion energy, $\Delta = \log ( E_{\rm obs}/E_{\infty} )$.  For
compactness and readability in the following equations, we define
$\epsilon = \log(E/10^{51}\, {\rm erg})$.   The
posterior distribution for $\Delta$ is 
\begin{multline}
\label{eq:posterior}
P(\Delta | \{ \epsilon_{\rm obs,i} \} ) \propto \\ \prod_i \int
P(\epsilon_{{\rm obs},i} |
\epsilon_{\infty},\Delta) P(\epsilon_{\infty}|M) P(M) P(\Delta) \, d\epsilon_{\infty} dM \, .
\end{multline}
We assume a uniform prior for $\Delta$, $P(\Delta)$. 
This posterior distribution represents a hierarchical Bayesian
inference where $\epsilon_{\infty}$ and $M$ are intermediate nuisance
parameters.  The portion of the likelihood that is $P(\epsilon_{\infty}) =
\int P(\epsilon_{\infty}| M) P(M) \, dM$ has already been calculated and is
shown in Figure~\ref{fig:EsimPosterior}.   Since the distributions for
$P(\epsilon_{\infty})$ are essentially Gaussian, we use the
Gaussian approximation for $P(\epsilon_{\infty})$.  With this
approximation, the marginalization over the nuisance parameters in
eq.~(\ref{eq:posterior}) is analytic, and the likelihood is now given by
\begin{multline}
\mathcal{L}(\epsilon_{{\rm obs},i} | \epsilon_{\infty}, \Delta) = \\ 
\frac{1}{\sqrt{2 \pi (\sigma_{\rm obs}^2 + \sigma_{\infty}^2)}}
e^{-[\epsilon_{{\rm obs},i} - \mu_{\infty} - \Delta]^2/[2(\sigma^2_{{\rm obs},i}
	+ \sigma_{\infty}^2)]} \, ,
\end{multline}
where $\mu_{\infty}$ is the mode of $P(\epsilon_{\infty})$, and
$\sigma_{\infty}$ is the half width of the 68\% HDI.

\begin{figure}
%	% Allowable file formats are eps or ps if compiling using latex
%	% or pdf, png, jpg if compiling using pdflatex
	%\includegraphics[width=\columnwidth]{data/InferMissingEOnly.pdf}
	\includegraphics[width=\columnwidth]{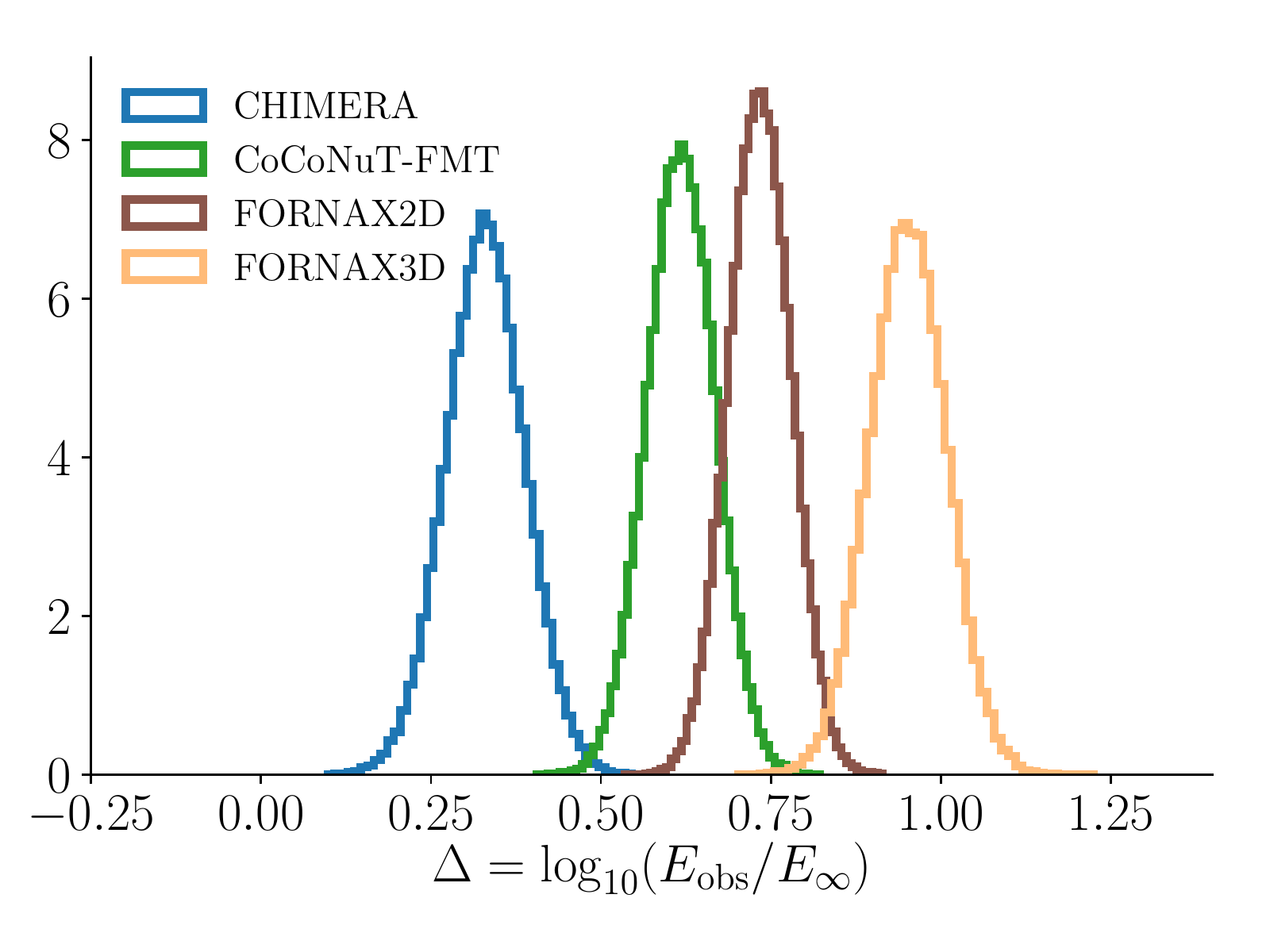}
    \caption{Posterior distributions comparing observational and
	  simulated explosion energies.  All three codes produce
	  simulation energies that are significantly lower than explosion
	  energies inferred from observations.  For CHIMERA,
	  $\Delta = \deltachimera$, for CoCoNuT-FMT $\Delta =
	  \deltacoconut$, for FORNAX2D,  $\Delta = \deltafornaxTwoD$, and
	  for FORNAX3D, $\Delta = \deltafornaxThreeD$.  The mean $\Delta$
	  for all simulations is $\sim$0.7; the variation for the simulations
	  ranges from 0.33 to 0.95 (width of $\sim$0.6).  The variation
	  among the simulations is as large as the mean offset.  While
	  there is a tension between the simulations and observations, the
	  large variation among simulations suggests that further
	  improvements to simulations could resolve this discrepancy.}
    \label{fig:MissingEnergy}
\end{figure}

Figure~\ref{fig:MissingEnergy} shows the inferred $\Delta$
distributions for the four simulation sets.  For CHIMERA, 
$\Delta = \deltachimera$;
for CoCoNuT-FMT, $\Delta = \deltacoconut$; 
for FORNAX2D $\Delta = \deltafornaxTwoD$;
for FORNAX3D $\Delta = \deltafornaxThreeD$.  All three simulations set
have more than 99.9\% of their distributions, $P(\Delta)$, greater
than zero.  Based upon the models and assumptions in this manuscript, all three
simulation sets have explosion energies that are significantly smaller
than observations.  In the best case (CHIMERA), the
simulated explosion energies are a factor 2 less energetic than the
observed energies.  In the worst case (FORNAX3D), the simulated
explosion energies are a factor of 10 less energetic.

\section{Discussion \& Conclusions}
\label{sec:Conclusion}

In general, we find that the explosion energies of multi-dimensional
simulations are significantly less energetic than the explosion
energies inferred from observations.  For this comparison, we require
the CCSN simulations and codes to have the following attributes; two-
or three-dimensional, neutrino transport that is a self-consistent approximation
of Boltzmann transport equations, transport is multi-angle and
multi-species, general relativity or some post-Newtonian
approximation, and positive explosion energies, approaching asymptotic
values.  The codes that satisfy these requirements are CHIMERA, CoCoNuT-FMT, and FORNAX.
For the observational explosion energies, we use the results of
\citet{pejcha2015} and \citet{muellerprieto2017}, who infer the explosion energies, uncertainties,
and correlations for 38 type IIP supernovae.  We infer a mean observational
explosion energy of $\mu_{\rm obs} = \muobs $ in units of
$\log_{10}(E_{\rm obs}/10^{51}\, {\rm erg})$; the width of the
distribution is $\sigma_{\rm obs} = \sigmaobs$.  To compare the
observations and simulations, we infer the ratio of
observed-to-simulated explosion energies, $\Delta = \log_{10}(E_{\rm obs}/E_{\infty})$, where $E_{\infty}$ is our estimate of the
simulation energy extrapolated to infinite time.  
For CHIMERA, $\Delta = \deltachimera$,
for CoCoNuT-FMT, $\Delta = \deltacoconut$,
for FORNAX2D, $\Delta = \deltafornaxTwoD$, and
for FORNAX3D, $\Delta = \deltafornaxThreeD$.  Overall, all simulation
sets are less energetic than the explosion energies inferred from
observations.  

This result suggests a tension between the simulations and
observations.  However, it does not yet rule out the standard neutrino
mechanism.  Of the four simulation sets, the mean offset is $\Delta
\approx 0.7$, but the range goes from 0.33 to 0.95, and the width of
this range is $\sim$0.6.  In other words, the variance among different
simulation sets is of order the average offset.  Given this, it is
plausible that future improvements to the simulations might
resolve the current discrepancy.

In this analysis, we identify several biases in the simulation sets.  
Core-collapse simulations are computationally expensive, and few of
the results include large systematic studies of the full range of
progenitors.  Furthermore, most simulations terminate well before an asymptotic
explosion energy.  

To mitigate for these biases in the simulation sets, we model the distribution of explosion energies from
simulations.  The FORNAX2D results exhibit the largest sample, so we use
their results to infer an explosion energy vs. progenitor mass
relationship.  Our inference shows that $E_{\infty} \propto M^{1-2}$;
the simulation explosion energies are proportional to progenitor mass
with a power ranging from linear to quadratic.  To infer the simulation
explosion energy after infinite time, we fit a simple model, 
$E_{\rm sim}(t) = E_{\infty} - A/t$, that is motivated by a simple
neutrino-powered explosion.  Then we assume that all models between
the 7.4 M$_{\odot}$ and 25 M$_{\odot}$ explode.  This range includes
the minimum mass for CCSN explosions \citep{diaz-rodriguez2018} and a
rough estimate for the maximum progenitor for SN IIP
\citep{smartt2015,davies2018}.

Better estimates for the simulated explosion energy distribution
will require more systematic explorations between 7.4 and 25
M$_{\odot}$, and better estimates will require simulations that
terminate later.  Based upon our simple model, simulations must
evolve roughly 0.5 to 2 seconds past the time of positive explosion energies to
reach at least 90\% of the asymptotic explosion energy.

There are other potential biases in the simulations which are either
difficult to quantify in this study or have yet to be identified at
all.  For example, resolution of the grid may impact whether CCSN
simulations have converged.  Recently, \citet{melson2019} explore how
resolution affects turbulence in simplified three-dimensional CCSN
simulations, but these explorations do not address how resolution
affects the explosion energies.  A major difference among the codes is
the treatment of neutrino transport.  There are many approximations
and choices in the transport: ray-by-ray vs. multi-angle, energy
groups, moments vs. short characteristics, moment closures,
velocity-dependent terms, gravitational redshift, scattering
opacities, correlated opacities, etc.  In fact, there are more
different choices in the neutrino transport treatment than there are
codes.  Yet, it is unclear how these differences impact the explosion
energies.

Progenitor perturbations due to O-shell burning may be important in
reducing the tension between simulations and observations.  The majority of simulations that are
available do not include progenitor perturbations, but the one model
that does, offers some tantalizing clues.  The 18ProgConv model of the
CoCoNuT-FMT set has the largest explosion energy of that set.  If we
include this one model, then the
discrepancy reported for CoCoNuT-FMT does not actually change much.  The lack
of change is
because this model has a relatively high mass.  The explosion energy vs.
progenitor mass correlation already indicates that the higher masses
explode with higher energy.  Including 18ProgConv in the fit only
steepens the dependence a little.  In addition, the highest
masses are more rare and provide little weighting to the final
explosion energy distribution.  So, this one perturbation-aided
explosion does not change $\Delta$ for CoCoNuT-FMT.  

However, it is not clear what progenitor perturbations would do for the lower mass
progenitors.  \citet{mueller2017} report a difference in explosion
time depending upon the size of the progenitor perturbations.  They
simulated three 18-M$_{\odot}$ progenitors models: one with no O-shell
perturbations, one with a convective mach number of 0.04, and one with
a convective mach number of 0.1.  The first did not explode by 650 ms
after bounce, the second exploded at around 500 ms, and the largest
perturbations exploded at 300 ms.  They only show the inferred
explosion energy for the largest perturbations, so we do not have a
quantitative measure of how perturbations affect the explosion energy.
None the less, an earlier explosion might lead to higher explosion
energies.  To test whether progenitor perturbations affect the
explosion energies, we recommend that simulators perform a systematic
study of progenitor perturbations for a wide range of masses.

Throughout this manuscript, we are careful to note that the explosion
energies that represent the observations are not observations in
themselves, but are inferences based upon observations.  As such,
the ``observed'' explosion energies are also subject to biases.
The observational explosion energy set relies on modeling
type IIP light curves and spectra.  Currently, there are two general
approaches.  One is to use one-dimensional radiation hydrodynamic
simulations to model the light curve and spectra \citep{kasen2009}.  The second is to
use fitting formulae to infer explosion energies,
etc \citep{pejcha2015,mueller2017}.  The modeling efforts are potentially more
accurate, but the fitting formulae are more amenable to statistical
inference and better uncertainty estimates.  For this preliminary
analysis, we find the distribution of explosion energies from light
curve modeling and fitting formulae to be consistent.  Therefore, we
use the statistical inference results from the fitting formulae.  A more accurate
and precise estimate would be to use the light curve models in a
statistical inference framework.

A significant source of systematic uncertainty in modeling the photospheric
properties is the zero point in the fitting formulae.
\citet{goldberg2019} summarize the zero points for several studies \citep{popov1993,kasen2009,sukhbold2016,goldberg2019};
see the discussion just after eq.~(7) in their manuscript.  They find
that the systematic uncertainty in the luminosity zero points is about
0.09 in log base 10.  This translates to a systematic uncertainty in
the log of the explosion energy of 0.1.  While this is not enough to
completely account for the discrepancy between the simulations and
observations, it is of the same order.
In addition, \citet{dessart2019} and \citet{goldberg2019} caution
that there are significant degeneracies among the explosion parameters:
nickel mass, ejecta mass, explosion energy, and progenitor radius.  In
fact, \citet{goldberg2019} argue that one other observational parameter besides
$M_V$, velocity, and $t_p$ is required to break this significant degeneracy.

Another potential source of bias for the observational set is the sample
of SNe.  At the moment, most modelers infer explosion energies by modeling
light curves of type IIP SNe.
For this analysis, we assume that all progenitors between 7.4 and 25
M$_{\odot}$ explode as type IIP SNe.  However, it is not clear what
fraction of this mass range corresponds to IIL or even Ib/Ic.  The
recent progenitor mass inferences of 25 historic SNe
\citep{williams2018} suggest that at least some fraction of this range
do correspond to these other SN types.  SN surveys suggest that SN IIP
are only $48.2^{+5.7}_{-5.6}$\% of all CCSNe \citep{smith2011a}.  At the
moment, it is not clear if this fraction is a result of a mass
dependence or binary evolution.  Whatever the case may be, there is a
clear bias in the observed explosion energies for a sub sample of
CCSNe.  One strategy to mitigate against this potential source of bias
would be to model the explosion energies of all SN types within a
volume-limited sample.  To do this, light curve models must include
the other SN types, not just SN IIP. 

In summary, we find that the explosion energies of multi-dimensional
CCSN simulations are significantly lower than the energies inferred
from observations.  Depending upon the simulation set, they are less
energetic by a factor of 2 to 10.  This suggests that either something is missing in CCSN simulations or there
are biases in our comparison.  We identify several sources of bias for
both the simulated and observed sets.  In this preliminary analysis,
we model some of these biases, but we recommend several ways to reduce
the impact of these biases in the future.  Given these biases, it is
probably premature to make any conclusions about the fidelity of CCSN codes.
Rather, the primary conclusion is that the current simulation and
observational sets are inconsistent, all suffer from biases, and the
path toward constraining CCSN theory requires careful consideration of
the biases in both.

%\begin{table}
%	\centering
%	\caption{Explosion energies for 2D and 3D CCSN simulations.}
%	\label{tab:explosionenergies}
%	\begin{tabular}{lr} % four columns, alignment for each
%		\hline
%		Progenitor & $E_{\rm expl}$ $10^{50}$ \\
%		\hline
%		\multicolumn{2}{c}{3D Sim. in \citet{melson2015}}\\
%		\hline
%		9.6 & 0.98\\
%		\hline
%		\multicolumn{2}{c}{3D Sim. in \citet{mueller2015}}\\
%		\hline
%		11.2 & 1.28\\
%		\hline
%		\multicolumn{2}{c}{3D Sim. in \citet{mueller2018}}\\
%		\hline
%		z9.6 & 1.32\\
%		s11.8 & 1.99\\
%		z12 & 4.1\\
%		s12.5 & 1.56\\
%		\hline
%		\multicolumn{2}{c}{2D Sim. in \citet{bruenn2016}}\\
%		\hline
%		B12-WH07 & 3.13\\
%		B15-WH07 & 8.84\\
%		B20-WH07 & 3.75\\
%		B25-WH07 & 7.02\\
%		\hline
%		\multicolumn{2}{c}{2D Sim. in \citet{vartanyan2018a}}\\
%		\hline
%		12\_WH07\_SFHo & 0.00\\
%		13\_WH07\_SFHo & 0.00\\
%		15\_WH07\_SFHo & 0.00\\
%		16\_WH07\_SFHo & 1.64\\
%		17\_WH07\_SFHo & 2.90\\
%		19\_WH07\_SFHo & 2.39\\
%		20\_WH07\_SFHo & 0.00\\
%		21\_WH07\_SFHo & 0.00\\
%		25\_WH07\_SFHo & 0.00\\
%		\hline
%		\multicolumn{2}{c}{2D Sim. in \citet{Radice2017}}\\
%		\hline
%		n8.8 & 1.78\\
%		u8.1 & 1.03\\
%		z9.6 & 1.22\\
%		9.0 & 0.63\\
%		10.0 & 0.12\\
%		11.0 & 1.35\\
%		\hline
%		\multicolumn{2}{c}{3D Sim. in \citet{vartanyan2019}}\\
%		\hline
%		16\_WH07\_SFHo & -0.79 (1.64 2D)\\
%		\hline
%	\end{tabular}
%\end{table}

\section*{Acknowledgements}

The authors thank Bernhard M{\"u}ller, Hans-Thomas Janka, Ondrej
Pejcha, and Jose Prieto for their careful scrutiny and thoughtful suggestions.
This work has been assigned an LANL document release number LA-UR-19-23383.

%%%%%%%%%%%%%%%%%%%%%%%%%%%%%%%%%%%%%%%%%%%%%%%%%%



\bibliographystyle{mnras}
%\bibliography{jeremiah.bib} % if your bibtex file is called example.bib

% Don't change these lines
\bsp	% typesetting comment
\label{lastpage}
\end{document}